\documentclass{aastex}
\usepackage{emulateapj5}

\newcommand{\ml}{$M/L$}
\newcommand{\msun}{M$_\sun$}
\newcommand{\lsun}{L$_\sun$}
\newcommand{\up}{$u^\prime$}
\newcommand{\gp}{$g^\prime$}
\newcommand{\rp}{$r^\prime$}
\newcommand{\ip}{$i^\prime$}
\newcommand{\zp}{$z^\prime$}
\newcommand{\us}{$u^\ast$}
\newcommand{\gs}{$g^\ast$}
\newcommand{\rs}{$r^\ast$}
\newcommand{\is}{$i^\ast$}
\newcommand{\zs}{$z^\ast$}
\newcommand{\hinv}{$h^{-1}\ $}
\newcommand{\om}{$\Omega_m$}
\newcommand{\ol}{$\Omega_{\Lambda}$}
\newcommand{\map}{$M_{260}$}

\newcommand{\mlap}{$M_{260}/L_{central}$}

\slugcomment{Submitted to Astrophysical Journal: Last Modified
13-June-2001. Version 1.5}

\shortauthors{McKay et al.}
\shorttitle{Galaxy Mass/Luminosity Scaling Laws}

\begin{document}


\title{Galaxy Mass and Luminosity Scaling Laws Determined 
by Weak Gravitational Lensing}


\author{
Timothy A. McKay\altaffilmark{1},
Erin Scott Sheldon\altaffilmark{1},
Judith Racusin\altaffilmark{1},
Philippe Fischer\altaffilmark{2},
Uros Seljak\altaffilmark{3},
Albert Stebbins\altaffilmark{4},
David Johnston\altaffilmark{5},
Joshua A. Frieman\altaffilmark{4,5},
Neta A. Bahcall\altaffilmark{3},
J. Brinkmann\altaffilmark{6},
Istv\'an Csabai\altaffilmark{7,8},
Masataka Fukugita\altaffilmark{9},
G. S. Hennessy\altaffilmark{10},
Robert B. Hindsley\altaffilmark{11},
\v{Z}eljko Ivezi\'{c}\altaffilmark{3},
D.Q. Lamb\altaffilmark{5},
Jon Loveday\altaffilmark{12},
Robert H. Lupton\altaffilmark{3},
Jeffrey A. Munn\altaffilmark{13},
R. C. Nichol\altaffilmark{14},
Jeffrey R. Pier\altaffilmark{13},
Donald G. York\altaffilmark{5,15}
}

\altaffiltext{1}{University of Michigan, Department of Physics, 500 East 
University, Ann Arbor, MI 48109}
\altaffiltext{2}{Deptartment of Astronomy, Univ. of Toronto, Toronto, ONT, 
M5S, 3H8, Canada}
\altaffiltext{3}{Princeton University Observatory, Princeton, NJ 08544}
\altaffiltext{4}{Fermi National Accelerator Laboratory, P.O. Box 500, 
Batavia, IL 60510}
\altaffiltext{5}{The University of Chicago, Department of Astronomy and 
Astrophysics, 5640 S. Ellis Ave., Chicago, IL 60637}
\altaffiltext{6}{Apache Point Observatory, P.O. Box 59, Sunspot, NM 88349-0059}
\altaffiltext{7}{Department of Physics and Astronomy, The Johns Hopkins 
University, 3701 San Martin Drive, Baltimore, MD 21218}
\altaffiltext{8}{Department of Physics of Complex Systems, 
E\"otv\"os University, P\'azm\'any P\'eter s\'et\'any 1}
\altaffiltext{9}{University of Tokyo, Institute for Cosmic Ray Research, 
Kashiwa, 2778582, Japan}
\altaffiltext{10}{U.S. Naval Observatory, 3450 Massachusetts Ave., NW, 
Washington, DC  20392-5420}
\altaffiltext{11}{Remote Sensing Division, Code 7215, Naval Research 
Laboratory, 4555 Overlook Ave. SW, Washington, DC 20375}
\altaffiltext{12}{Astronomy Centre, University of Sussex, Falmer, 
Brighton BN1 9QJ, United Kingdom}
\altaffiltext{13}{U.S. Naval Observatory, Flagstaff Station, P.O. 
Box 1149, Flagstaff, AZ  86002-1149}
\altaffiltext{14}{Dept. of Physics, Carnegie Mellon University, 5000 
Forbes Ave., Pittsburgh, PA-15232}
\altaffiltext{15}{The University of Chicago, Enrico Fermi Institute, 
5640 S. Ellis Ave., Chicago, IL 60637}


\begin{abstract}
We present new measurements of scaling laws relating the luminosity 
of galaxies to the amplitude and shape of their dark matter halos. 
Early imaging and spectroscopic 
data from the Sloan Digital Sky Survey are used to make 
weak lensing measurements of the surface mass density contrast
$\Delta\Sigma_+$ around classes of lens objects. This
surface mass density contrast as a function of radius is a measure of 
the galaxy-mass correlation function (GMCF). Because spectroscopic
redshifts are available for all lens objects, the mass and distance
scales are well constrained.
The GMCF measured around $\sim$31,000 lenses is well fit by a power law
 of the form 
$\Delta\Sigma_{+} = (2.5^{+0.7}_{-0.6}) 
	(R/1 $Mpc$)^{-0.8\pm0.2} h $M$_{\sun} $pc$^{-2}$.
We compare this GMCF to galaxy luminosity, type, and environment, and 
find that it varies strongly with all three. We quantify these variations
by comparing the normalization of a fit to the inner 260 \hinv kpc
($M_{260}$) to the galaxy luminosity. While $M_{260}$ is 
not strongly related to luminosity in 
bluest band (\up), there is a simple, linear relation 
between $M_{260}$ and luminosity in redder bands (\gp, \rp, \ip, and \zp).
We test the universality of these mass-to-light scalings by independently 
measuring them for
spiral and elliptical galaxies, and for galaxies in a variety of 
environments. We find remarkable consistency in these determinations in 
the red bands, especially \ip\ and \zp. 
This consistency across a wide range of systems suggests
that the measured scaling represents an excellent cosmic average, and that
the integrated star formation history of galaxies is strongly related
to the dark matter environments in which they form. Future studies of 
galaxy mass and its relation to luminosity should concentrate on 
luminosities measured in red bands.
\end{abstract}



\section{Introduction}

\subsection{The galaxy-mass correlation function} \label{introduction}

The relationship between the luminous matter which we observe in the Universe 
and the dark matter which dominates its dynamical evolution is elusive. 
Models of hierarchical structure formation based on N-body simulations now 
provide a relatively complete picture of the formation, evolution, and 
clustering properties of dark matter halos. Unfortunately, experimental
determination of structure relies on observations of luminous galaxies.
The formation of luminous galaxies involves a variety of complex physical
phenomena; gas physics, star formation, and the feedback mechanisms
into the interstellar medium. These processes are currently too complex 
for a direct simulation. As a result, the relationship between 
luminous galaxies and the dark matter environment in which they form
is poorly determined. This
uncertainty seriously limits our 
ability to directly compare cosmological observations of galaxies
to N-body simulations.
Measurements which connect the mass distribution so carefully studied in
N-body simulations to the luminous galaxies which we observe play a crucial
role in understanding the formation of structure in the universe.

The connection between luminous galaxies and the dark matter structures
in which they reside has traditionally been approached by attempting to
measure the masses of galaxies as discrete objects. As our 
understanding of the formation of cosmic structure has evolved, 
it has become clear that the dark halos containing galaxies are not
discrete structures, clearly separated from one another. N-body 
simulations reveal instead a continuous matter field, with structure 
gradually accreting on all scales. Galaxies are expected to form in 
collapsed halos, which merge with each other into larger halos. These 
halos typically extend out to large radii without any apparent cut-off 
in the density profile.
At each level one expects to find several galaxies inside the halo of 
a given mass, but some of these may be too faint to enter into the
observational sample.
In this picture the concept of `the' 
mass of a galaxy may be poorly defined. 
While it is still possible to discuss, in an average sense, the
mass profile of a galaxy, or to measure the mass within a fixed aperture,
it may be inappropriate to speak of a galaxy's total mass.

A more appropriate way to describe the 
relationship between luminous matter and mass is through correlation measures.
The full statistical relationship between galaxies and dark matter can be 
expressed as a Galaxy-Mass Correlation Function (GMCF). Given a set 
of points marking the locations of galaxies, a GMCF measures the 
extent to which mass is clustered around these points. A simple version
of the GMCF would describe the average projected mass density 
interior to a radius R from a 
galaxy location. If there were no correlation
between the locations of galaxies and mass, this GMCF would be flat. If
galaxies were isolated point objects with discrete and equal 
masses, the GMCF would fall
off as $R^{-2}$. If galaxies are embedded in isolated isothermal dark halos 
with equal mass, 
the GMCF will fall off as $R^{-1}$.

If the luminous properties of galaxies (luminosity, morphology, stellar 
population, etc.) are affected by the dark matter environment in which 
they form, we expect this GMCF to vary as we select different classes of 
galaxies. In this sense, crucial information about the connection 
between luminous and dark matter is encoded in the scaling laws 
describing changes in the amplitude and shape of the 
GMCF with galaxy properties. 

In this work we measure something closely related to the naive 
GMCF using weak gravitational lensing; the projected surface mass density 
contrast. 
It is the difference between the surface density interior to a projected
radius R and
the surface density at the radius R. This surface mass density 
contrast function is what
we refer to in this paper as the galaxy-mass correlation function (GMCF). 
We will measure the way in which 
the amplitude and shape of this GMCF scales with lens luminosity, 
morphology, and local 
environment to better understand how luminous and dark matter in the universe
are related. These measurements will be made in all five SDSS colors, 
allowing us to probe the relationship between global galaxy 
properties and various stellar populations.

A traditional way of expressing the relations between mass and light 
is to compare an aperture mass and luminosity in a region studied. 
This mass-to-light ratio \ml, measured in solar units 
\msun/\lsun, provides a 
convenient expression of the relative importance of dark and luminous
matter in the region studied. It can be dependent on 
wavelength, and hence is usually denoted by an observed pass-band; for 
example \ml$_{g^{\prime}}$ is the mass-to-light ratio, in solar units, measured in
the SDSS \gp\ band. When we speak of 
{\it a} mass-to-light ratio we implicitly 
assume a linear relationship between mass and light. In general we 
expect \ml\ to be a function, which may depend on 
scale, luminosity, galaxy type, and environment.

\subsection{Methods of measuring galaxy masses}

The quantity and distribution of mass in the universe 
has usually been determined
by dynamical means. Motions of luminous test bodies are used to
constrain mass distributions. Test masses are, of course, primarily sensitive
to mass interior to the radius of their orbits.
Within galaxies, the motions of stars and
gas give strong evidence for dark components of disks and for extremely
massive, extended halos of dark matter \citep{sac95}. On halo scales 
satellite galaxies can be used to probe the matter
distribution. Studies of satellites give strong, but model dependent, 
suggestions that dark matter halos extend far beyond the luminous cores of 
galaxies \citep{zar97,zar99}, merging smoothly into the halos of their
neighbors.

On larger scales, the presence of mass can be inferred from the motions of
galaxies in groups \citep{gir00}, clusters \citep{car97},
and beyond \citep{wil00}. In relaxed groups and clusters mass is 
also revealed by the emission of thermal x-rays from a hot intergalactic 
medium. This x-ray emission can be used to place strong constraints on
group and cluster masses on these large scales \citep{evr96,hra00}.

Dynamical measures of mass are extremely powerful. They have revealed the
dominant role of dark matter in the universe, and given us important hints
about its distribution. They have shown \citep{bah00} that \ml, measured
around galaxy centers, increases smoothly from small (10 \hinv kpc) to
large (250 \hinv kpc) scales. At larger scales \ml\ is more or less constant.
This increase with scale is not surprising, it reflects how
baryonic material cools and collapses to the center of dark
matter halos. The baryons collapse into a rotationally 
supported disk, with its scale determined by the angular momentum of 
baryons. Assuming the angular momentum is not transferred from
dark matter to baryons and that its initial value is appropriate for CDM models
one finds the disk scale length at least a factor of 10 smaller than the extent 
of the dark matter halo. 
As a result, most of the light is found in the centers of 
dark matter potential wells, while most of the mass remains in an 
extended halo.

A limitation of the dynamical approach is that it 
requires the presence of luminous tracers. In addition, the motions of 
test particles are very little affected by the presence of matter outside
the radius of their orbits. As a result, dynamical measures have
primarily measured the mass of the inner parts ($<$30 \hinv kpc) 
of galaxy halos.
Naive comparison of dynamical masses measured on various scales has led
to considerable confusion \citep{koc96,zar99}. 
More important, the modeling which is required
to make dynamical measurements of mass becomes increasingly uncertain on large
scales, where dynamical time scales can exceed the Hubble time. For these
reasons, it is essential to pursue alternative approaches.

Einstein's theory of general relativity predicts that the path of a light
ray will be deflected as it passes through an inhomogeneous mass 
distribution. The effects created by these deflections, referred to as 
gravitational lensing, provide an alternate approach to determining the
distribution of mass in the Universe. The gravitational lensing effect
created by an astrophysical object depends on both the density contrast 
within it and the geometry of the lensing system. It is completely
independent of the dynamical state or nature of the matter in the 
lens.

Gravitational lensing is detected by measuring the effect of nearby `lens'
objects on the observed shapes of more distant `source' galaxies.
The effects caused by high density lenses can be spectacular, including both 
multiply imaged sources \citep{wal79} and highly distorted giant arcs
\citep{lyn86}. In these cases of dramatic distortion, the lensing masses of
individual objects can be traced in detail. Unfortunately such strong
lensing occurs only in very rare, high density contrast regions. 

The density contrast in the outer regions of galaxies and clusters is very 
small. As a result, the lensing distortions they introduce can be tiny, 
often smaller than 1\%. The source
galaxies we observe have unknown intrinsic shapes and these weak 
lensing effects can only be observed statistically \citep{tys84,web85,tys90}. 
By measuring the average 
distortion in the shapes of a large number of background galaxies, we can 
determine the effect of a lens. This weak lensing technique
depends on the assumption that alignments of source galaxies are produced
only by intervening lenses \citep{tys85}. 
Source galaxies seen as close to a lens in angle are usually separated 
from the lens by large distances along the line of sight,
so this assumption is not unreasonable.

In the case of galaxies, the effect of a single lens is too
small to be measured accurately using the limited number of source objects 
available behind it. In this case we can still measure the $\it average$
lensing effect of a set of lenses by `stacking' the lens objects. This is
fundamentally the same as measuring the objects individually (at very low 
signal-to-noise) and combining the measurements. This technique
of stacking lens objects was first performed for galaxy-galaxy
lensing studies \citep{tys84,bra96,fis00,smi00,wilson00}. 
Its use has now been extended to the statistical study of galaxy 
groups \citep{hoe01} and clusters \citep{she01}.

These statistical measures have often been considered as measurements
of the average lensing mass of a class of galaxies. This interpretation
is complicated by the intrinsic clustering of galaxies 
\citep{fis00,sel00} and by the fact that galaxies can be found in a range
of halo masses. On larger scales groups and clusters can dominate the signal
and since there are typically many galaxies inside such halos, only one 
of which can be at its center, one has to be careful with this interpretation.
Another way to say this is that
lensing masses measured on $\geq$ 100 kpc scales 
include contributions from both
the central galaxy being studied and its neighbors, all of which could be 
embedded in a larger mass concentration or could be clustered in the field. 

Interpretation of
lensing measurements is simplified by recognizing that
traditional galaxy-galaxy lensing studies really measure the correlation 
between a set of points (galaxy locations) and the mass which surrounds
them. This galaxy-mass correlation function (GMCF) is very well defined, 
and it can be directly compared to the results of N-body simulations or 
to other theoretical models just like the galaxy correlation function.
In this view the correlation of mass with a galaxy and the correlation
of galaxies with one another can be viewed as related, but not 
identical, clustering statistics, both of which in turn are related, but 
not identical, to dark matter clustering. This view is most useful on 
large scales, where the constant bias assumption is most reasonable. 
Here we will for the most part stick to the classical interpretation 
of galaxy-galaxy lensing as measuring dark matter halos around 
galaxies, but we will also comment on what this really means in 
light of more realistic models.

In this work we use gravitational lensing techniques to measure the 
galaxy-mass correlation function for a large magnitude limited sample of 
lens galaxies. We combine the derived GMCF with measurements of the 
optical luminosity associated with these objects in five optical pass-bands.
These measurements allow us to study the relation between
mass and light for galaxies covering a wide range in luminosity, 
galaxy type, and local environment. 

This work expands on our initial measurements of the GMCF \citep{fis00} in
two crucial ways. First, all lens objects used here have spectroscopic 
redshifts. This allows us to properly combine all lenses, and to place our
measurements on a solid physical scale. Second, since the distances to all
lenses are known, it is possible to confidently relate the physical properties
of the lenses to the lensing signals which they induce.

In \S \ref{observations} we describe the data on which this study is 
built and in \S \ref{lensmeasure} we
describe our gravitational lensing mass reconstructions.
We present measurements of the GMCF in \S \ref{galmasscorr}, and probe its
dependence on luminosity, galaxy type, and environment.
In \S \ref{m2l} we
compare aperture mass and luminosity measurements to study the relationship
between \ml\ and various galaxy properties.
We discuss the results in \S \ref{discussion}. We conclude in \S
\ref{future} with a description of future extensions 
of these measurements, both to higher accuracy, and to new approaches. 
Throughout this work we use H$_0$ = 100 h km s$^{-1}$ Mpc$^{-1}$, 
$\Omega_m$ =0.3, and $\Omega_{\Lambda}$ = 0.7.

\section{Observations} \label{observations}

The analyses reported in this work are based on observations obtained by 
the Sloan Digital Sky Survey\footnote{www.sdss.org}. The SDSS includes both 
imaging and spectroscopic surveys \citep{yor00} of approximately 10$^4$ square
degrees in the North Galactic Cap. The SDSS telescope \citep{sie01} is a
dedicated 2.5 m f$\backslash$5 survey telescope with a flat, 
essentially undistorted
3\degr\ field of view. Two instruments, a large imaging camera \citep{gun98}
and a 640 fiber plugger plate 
multifiber spectrograph \citep{uom01} alternate in the
focal plane of the telescope. Imaging data is taken only when conditions are
optimal (photometric skies, no moon, and seeing $<$1.5\arcsec).
Spectroscopic data are obtained during the remaining time.
Imaging data are acquired in drift scan mode, providing near-simulataneous 
observations in five bandpasses (\up, \gp, \rp, \ip, and \zp\ \citep{fuk96}) 
to a limiting magnitude of about \rs\ = 23.0.
The pixel scale for imaging observations is 0.396\arcsec\ per pixel, 
allowing Nyquist sampling for 1.0\arcsec seeing.
Spectroscopic targets selected from these images are observed with
the multifiber spectrograph beginning in the month following imaging
observations. The spectra obtained cover
a wavelength range from 390 to 910 nm with a resolving power of $\sim$2000.
Typical S/N for SDSS spectra is $>$100 per pixel. The SDSS will ultimately 
image more than 10$^8$ galaxies, and obtain spectra for more 
than 10$^6$ galaxies and 10$^5$ quasars. To date the SDSS has obtained 
imaging data for more than 3000 square degrees, and spectra for more
than 150,000 independent objects.

The data used in this analysis are drawn from SDSS commissioning runs.
Imaging data were obtained between Fall 1997 and Spring 2000. Seeing in
these early runs varied from 1.0\arcsec\ to 2.0\arcsec. 
Spectroscopic targets selected from this imaging data were observed with 
the SDSS spectrographs over a series of nights during 2000. The SDSS 
photometric system is not yet finally defined. To remind the reader of 
this, we refer to all measured magnitudes using the symbols \us, \gs\ etc., 
and to the passbands themselves as \up, \gp\ etc.

SDSS imaging data are reduced and calibrated by the SDSS photometric
(PHOTO: \citet{lup01}), astrometric (ASTROM: \citet{pier01}) and
calibration (MT: \citet{tuc01}) pipelines. These pipelines begin by cleaning
the raw images and obtaining astrometric solutions. They
extract objects from the images, deblending them where necessary,
measure a wide range of possible properties, 
and combine the five color data into an extensive 
photometric catalog of measured
object properties. In addition to the parameter catalog these pipelines 
provide small `atlas images' extracted from the full images around each 
objects. They also provide continuous, detailed information about the shape
and size of the PSF across the focal plane. We make use of the  
SDSS measured parameters, the atlas images of objects, and the PSF
information to perform the analyses presented below.

Spectroscopic targets are selected from the imaging data in three primary 
classes; the `main' galaxy sample, designed for traditional large scale 
structure studies out to z=0.2, a `luminous red galaxy' (LRG) sample, 
designed to probe peaks of the density field to z=0.5, and a quasar sample.
We concentrate in this work on the `main' galaxy sample, which accounts
for about 80\% of SDSS spectroscopy. These objects are selected by 
requiring that they be well resolved and have reddening corrected 
\citep{sch98} Petrosian magnitudes (see below) brighter than
\rs=17.6. Details of the spectroscopic target selection for the SDSS 
main galaxy sample will be discussed in \citet{str01}.

The spectra for these objects were reduced by the SDSS spectroscopic
pipeline (SPECTRO: \citet{fri01}). This pipeline extracts 1D spectra from
the 2D spectrograph images, calibrates them, and measures redshifts by 
cross-correlation with stellar templates. Redshifts are determined for 
about 98\% of all galaxy targets, with a typical velocity accuracy of 
$<$30 km/s. 

\subsection{Lens and source galaxy selection} \label{selection}

To conduct our lensing mass measurements we divide 
the observed sample of galaxies into a foreground lens sample and a 
background source sample.
For lens galaxies we select only SDSS main galaxy targets for which 
spectroscopy is complete. As a result, every lens object has a measured
spectroscopic redshift. This is essential for the accuracy of our 
lensing analysis. It 
allows us to tightly constrain the geometry of each lens and to
appropriately rescale all of our lensing measurements before we combine them. 

\begin{figure}[t]
\plotone{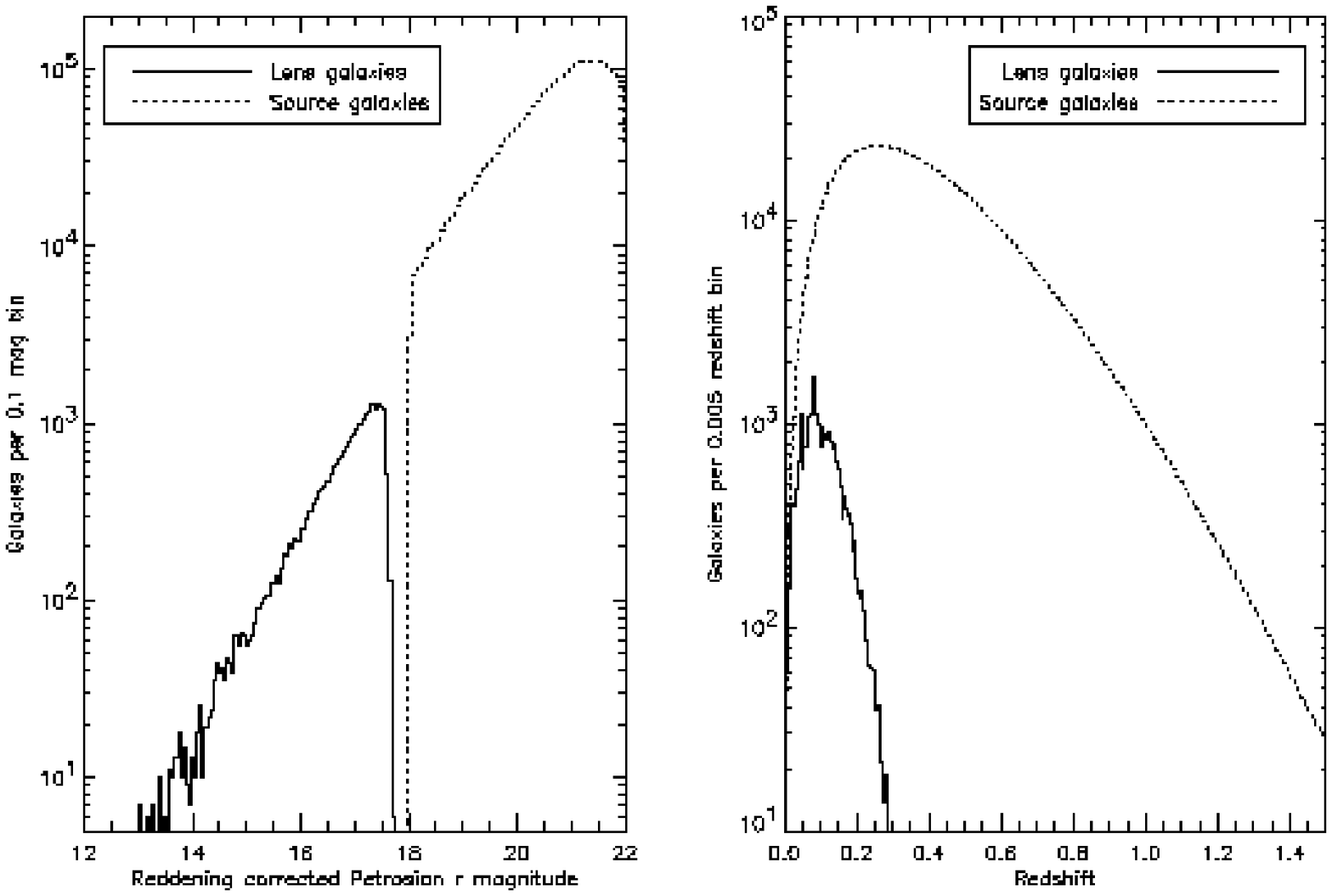}
\figcaption[f1_sm.ps]{Reddening corrected \rs\ magnitude 
distribution in bins of 0.1 mag (left) and redshift 
distribution in bins of .005
(right) for the lens and source galaxies used in this study. The redshift
distribution for the source galaxies is estimated by the methods described
in \S \ref{geometry}. The source galaxy magnitude distribution is higher
than one might expect by extrapolation of the lens distribution because we
we require completed SDSS spectroscopy to include a galaxy in the lens sample.
\label{lens_source_sample_plots}}
\end{figure}

The selection of galaxies used here is very similar to
that used in the first determination of the SDSS luminosity function
described in \citet{bla01}. The magnitude and redshift
distributions for this lens sample are shown in Figure 
\ref{lens_source_sample_plots}.
The photometric accuracy of this bright sample is limited solely by systematic 
calibration uncertainty of $\sim$3\%. The total number of lens galaxies 
available for these studies is 34693. 

Because of the physical size of the fiber holes in the spectrograph, it is
impossible to obtain spectra for two galaxies closer than 55\arcsec\
within one spectroscopic plate. Spectra are obtained for some of these
galaxies by overlapping spectroscopic plates, but a small fraction
($<5$\%) remain unobserved. While these galaxies are not analyzed as lenses,
we do keep track of them in the studies of lens angular clustering 
described below.

The source galaxy sample is assembled with two requirements in mind. First, 
source
objects should be behind the lens galaxies. Since we lack spectroscopic 
redshifts for these fainter objects, we begin by selecting all galaxies fainter
than \rs=18. To be useful for lensing, we must be 
able to accurately measure the shapes of source galaxies. It is 
impossible to accurately measure the shape of any object which is not well
resolved. The shape measured for objects which are smaller than the PSF 
is merely the shape of the PSF. Accurate shape measurement
also requires relatively high signal-to-noise (a $>$10$\sigma$ detection). 
As a result we select as source galaxies well resolved objects with 
\rs\ from 18.0 to 22.0. More details of the selection are given below. 

A total of 3,615,718 source galaxies are available
for these studies. Figure 
\ref{lens_source_sample_plots} illustrates the source magnitude 
and estimated redshift distribution. Estimation of the source redshift
distribution is described in \S \ref{geometry}.
Since the \gp, \rp, and \ip\ passbands are substantially more sensitive
than \up\ and \zp, we will conduct all our shape measurements of source
galaxies in these three central bands. The \up\ and \zp\ data for the 
relatively bright lens galaxies 
are, however, excellent. So we will use \up\ and \zp\ data in 
understanding lens properties.

\subsection{Lens luminosity measurements} \label{luminosity}

Luminosities for the lens galaxies are derived by combining the 
PHOTO measurements of light profiles in five colors with the redshifts
determined by SPECTRO. Since galaxies have a variety of luminosity profiles 
and lack well-defined boundaries, determining their total flux
is complex and model dependent. Traditional aperture and isophotal 
flux measurements are plagued by well known biases with redshift and 
galaxy type. As a result the SDSS has adopted a circular 
aperture flux measurement in which the radius of the aperture selected 
is adapted to the shape of each galaxy's light profile.

The basic measurement of galaxy luminosity for the SDSS is a 
\citet{pet76} magnitude. For the SDSS, the Petrosian radius is 
defined as the largest radius at which the local surface brightness 
is one eighth of the mean surface brightness interior to that
radius. The Petrosian flux is then the total flux within a 
circular aperture with radius two times the Petrosian radius. 
Technical details of this Petrosian flux measurement are presented in
\citet{lup01}. An extensive discussion of the fraction of total light 
measured for various model galaxy profiles is given in \citet{bla01}.
The SDSS Petrosian magnitude detects essentially all the light ($>$97\%)
from galaxies with exponential profiles, and more than 80\% of the light
from galaxies with de Vaucouleurs profiles.

Conversions from apparent magnitude m to absolute magnitude M depend on 
cosmology through the distance modulus DM(z) and on galaxy type through the
K-correction K(z). Absolute magnitudes are defined by the relation:
\begin{equation}
M = m - DM(z) - K(z)
\end{equation}
In this work we consider only a FRW cosmology in which \om=0.3 and \ol=0.7
with a Hubble constant H$_0=100 h$ km s$^{-1}$ Mpc$^{-1}$. The distance modulus
is then determined from formulae summarized in, for example, \citet{hog99}.
K-corrections are applied to correct for the difference between observed 
passbands and rest-frame passbands. Given the redshift and observed \gs-\rs\ 
color of each galaxy, we determine a best fit K-correction 
by interpolation from the tables 
supplied by \citet{fuk95}.

\begin{figure}[t]
\plotone{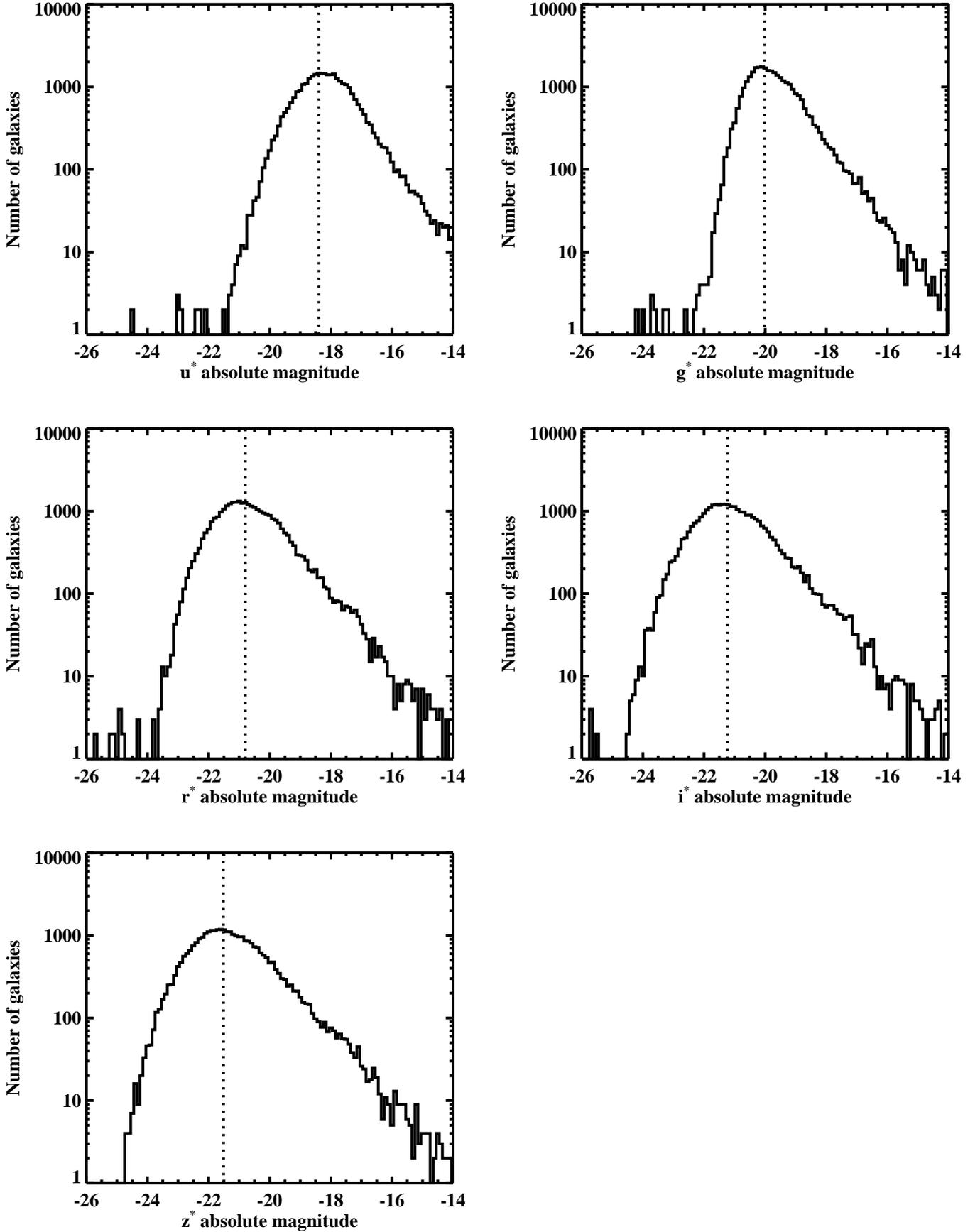}
\figcaption[f2.eps]{The absolute magnitude distributions
for the 34,693 lens objects are shown above. Magnitude bins are 0.1
magnitude. The dashed lines
mark the values of M$^{*}$ found in each band in \citet{bla01}.
\label{luminosity_distribution}}
\end{figure}

Luminosity distributions for all the lens galaxies in the sample are shown in
Figure \ref{luminosity_distribution}, along with the values of M$^{*}$ 
derived from SDSS measurements of the luminosity function \citep{bla01}.

\subsection{Lens galaxy classification} \label{classification}

As a part of this study, we will examine correlations between 
mass, luminosity, 
and lens galaxy type. To enable this we have conducted a simple
automatic classification of
all the lens galaxies in our sample. This classification is based on
a combination of SDSS PHOTO parameters and reanalysis of lens
atlas images. The goal of this classification is to place galaxies along a 
continuous sequence from early type `ellipticals' to late type `spirals'.

Previous studies of automatic galaxy classification \citep{abr94,mber00} have
shown that parameters such as concentration, color, and rotational asymmetry
correlate well with morphology or Hubble type. Concentration and color 
computed by PHOTO have been shown to correlate well with morphological type
by \citet{shi01} and \citet{isk01}. For this study we combine three 
parameters for galaxy classification: the \gs-\rs\ color, a measure 
of concentration, and a measure of asymmetry.

The \gs-\rs\ color has been shown by \citet{isk01} to correlate well with 
morphology for all galaxies within the redshift range of our lens sample.
For galaxies at higher redshift (beyond z$\sim$0.38), the 4000 \AA\ break
passes from the \gp\ to the \rp\ band, and classification must
consider the \rs-\is\ color.
We define a concentration parameter from PHOTO outputs by comparing 
$r_{50}^{p}(r^{\prime})$, the Petrosian half-light radius, and 
$r_{90}^{p}(r^{\prime})$, the radius at which 90\% of the \rp\ 
Petrosian flux is contained. Our concentration is the ratio of these two.

Unlike the rotational symmetry measurements used by \citet{abr94}, we employ
a bilateral asymmetry measurement. It is more sensitive than rotational 
symmetry to localized features because it does not dilute them as strongly.
This asymmetry measure is also model independent. Our asymmetry parameter is 
not measured in the standard PHOTO processing of SDSS data. To make the
measurement we refer to the `atlas images' of each object. After finding
the major axis of the light distribution of a galaxy, we subtract the galaxy
light on one side of the image from the light on the other. The 
asymmetry is then defined as:
\begin{equation}
\vspace{0.2cm}
{\textstyle Asymmetry} = \frac{\Sigma 
	{\textstyle Residuals\ from\ Subtraction}}
	{\Sigma {\textstyle Flux\ before\ Subtraction}}
\vspace{0.2cm}
\end{equation}
This parameter is small for galaxies with smooth, symmetric light profiles 
and large for galaxies which are lumpy or otherwise asymmetric. Because of 
the abundance of blue light emitted in star forming regions, we measure 
this asymmetry in the \gp\ band. 

\begin{figure}[t]
\plotone{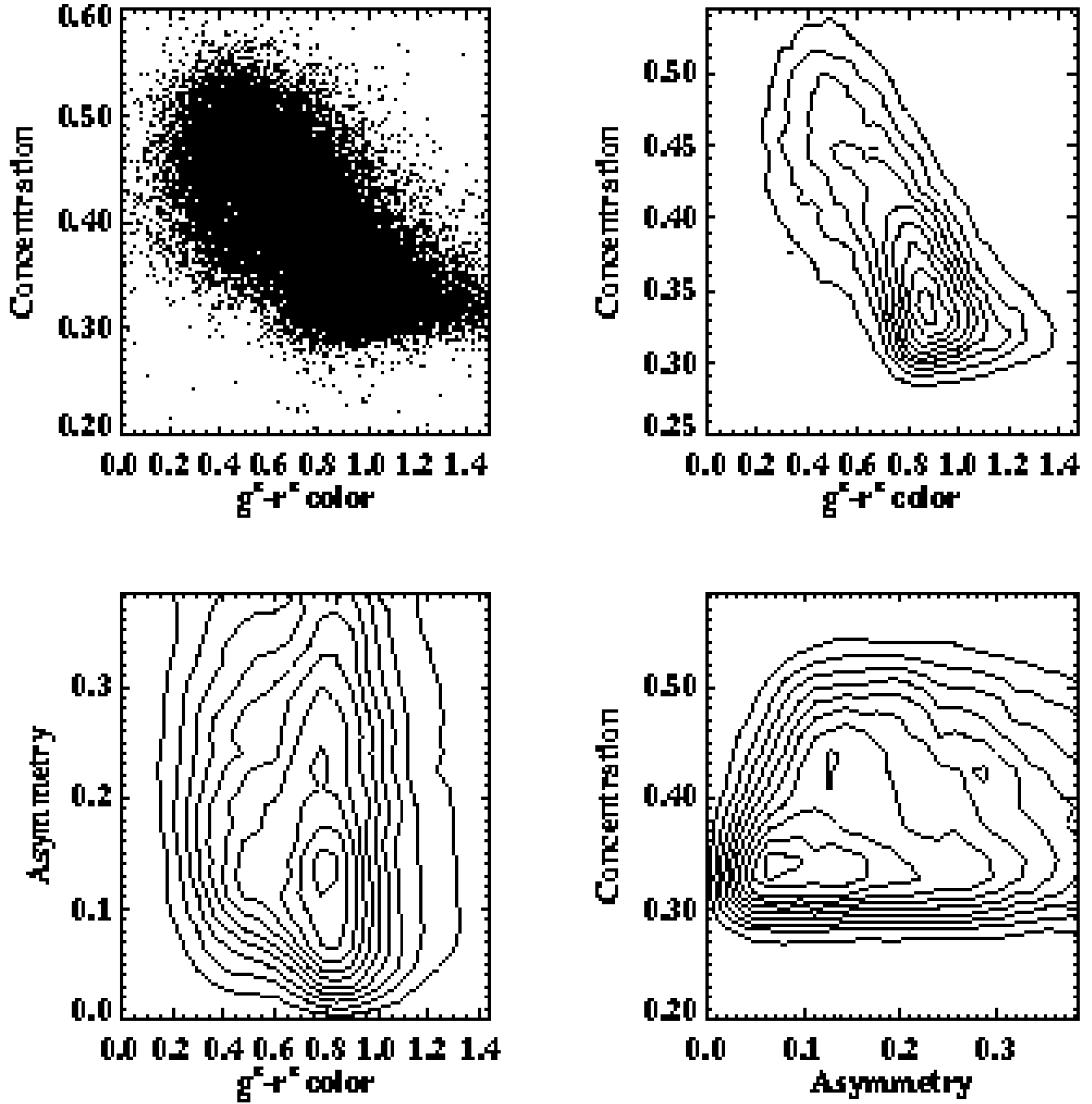}
\figcaption[f3_sm.ps]{This figure demonstrates the correlations 
among the three classification parameters described in \S 
\ref{classification}. On the upper left is a scatter plot of the
concentration vs. \gs-\rs\ color. This is included to give an idea of
the scatter of the points. The upper
right shows a contour plot of the distribution of galaxies in the 
concentration vs. \gs-\rs\ 
color plane. The lower left shows the relation between  
asymmetry and \gs-\rs\ color, and the lower right the relation between
concentration and asymmetry. 
\label{classification_1}}
\end{figure}

Correlations among these three classification
parameters are shown in Figure \ref{classification_1}. Concentration and
color contain most of the classification information, though the asymmetry 
parameter adds some discrimination among types.
These three classification parameters are scaled to their approximate
range and combined in quadrature to produce a single classification 
parameter which varies from zero to one. This classification is not intended
to place galaxies along a Hubble sequence, but merely to separate late
types from early types in a statistical way. Nevertheless, we find that
our classification correlates well with the visual 
classification of 456 bright SDSS galaxies reported by \citet{shi01}. 
We will later divide the lens catalog into subsets using this classification 
parameter.

\section{Lensing mass measurements} \label{lensmeasure}

When light from a source galaxy passes near a lens, its path is deflected. 
If the surface density of the lens is decreasing with projected
radius, the effect of lensing in the weak regime
is to stretch the images of background galaxies
in the tangential direction. There is a 
simple relation between the tangential shear and the mass density contrast:
\begin{equation} \label{tshear}
\gamma_+(R) = \overline{\kappa}(\leq R) - \overline{\kappa}(R)
\end{equation}
where $\gamma_+$ is the shear in the tangential direction, and 
$\kappa = \Sigma/\Sigma_{crit}$ is the surface density of the lens measured
in units of the critical density. The critical density is dependent
in an important way on the geometry of the lens-source system:
\begin{equation}
\Sigma_{crit} = \frac{c^2}{4\pi G} \frac{D_S}{D_L D_{LS}}
\end{equation} 
where $D_S$ and $D_L$ are the angular diameter distances to source and lens,
respectively, and $D_{LS}$ is the distance from source to the lens. 

If the 
lens geometry is known, measurements of the tangential shear can be converted
directly to measurements of surface mass density contrast in the lens:
\begin{equation} \label{densconteq}
\gamma_{+}(R)\Sigma_{crit} = \overline{\Sigma}(\leq R) 
	- \overline{\Sigma}(R) \equiv \Delta\Sigma_+
\end{equation} 
This measured mass density contrast can then be either integrated or, more
often, fit to a model profile, to constrain the mass of a lens object.
For the case of an isothermal mass profile ($\Sigma(R)\propto R^{-1}$), the
density contrast $\Delta\Sigma_+$ is equal to the density itself. This surface
mass density contrast is what we refer to in this work as the galaxy-mass
correlation function.

Two measurements are essential to lens reconstruction; the tangential
shear $\gamma_{+}(R)$ produced by the lens as a function of radius, and the 
geometry of the lens system as encoded in $\Sigma_{crit}$. 
Both are described in detail in what follows.

\subsection{Measurement of galaxy shapes} \label{shapemeasure}

The first step in measuring tangential shear is measurement of the source
galaxy shapes. To improve the S/N of these shape measurements beyond those
provided by the SDSS pipelines, we re-measure all object
shapes from the atlas images using an adaptively weighted moment 
scheme \citep{ber01}. We measure quadratic 
moments ($Q_{i,j}$) of the galaxies weighted by an elliptical Gaussian
matched in location, size, and orientation (by an iterative procedure)
to that of the object being measured:
\begin{equation}
Q_{i,j} = \frac{\sum_{k,l}I_{k,l}G_{k,l}x_{i}x_{j}}{\sum_{k,l}I_{k,l}G_{k,l}}
\end{equation}
where $I_{k,l}$ is the sky subtracted surface brightness of pixel (k,l), 
$G_{k,l}$ is the Gaussian weight evaluated at pixel (k,l), and $x_{i}$ is
the pixel offset from the image centroid. From these moments we define the 
ellipticity components, or polarization,  \citep{bla91}:
\begin{eqnarray}
e_1 = \frac{Q_{1,1} - Q_{2,2}}{Q_{1,1} + Q_{2,2}} \nonumber \\
e_2 = \frac{2Q_{1,2}}{Q_{1,1} + Q_{2,2}} 
\end{eqnarray}
These ellipticity components are related to the more familiar major axis (a),
minor axis (b), and position angle ($\beta$) as:
\begin{equation}
(e_1,e_2) = \frac{a^2 - b^2}{a^2 + b^2}(\cos{2\beta},\sin{2\beta})
\end{equation}

The measured shapes of these objects are then corrected to remove the effects
of seeing and an anisotropic PSF as described in \citet{fis00}. There are
two important effects in SDSS data. The isotropic part of the PSF
(mostly seeing) tends to circularize or dilute the observed shapes of galaxy 
images. The sensitivity of a galaxy to this circularization is called its
``smear polarizability'' S$_m$, which depends primarily on the ratio of
PSF size to galaxy size.
An object which is smaller in size than the PSF width has S$_m$ approaching
one.
It is impossible to measure ellipticity of such objects, as they are 
unresolved.
An object which is large compared to the PSF has S$_m$ approaching zero. 
We can measure the shapes of such large objects very accurately, 
independent of the size of the PSF. Though the smear polarizability is 
mostly due to object size, it is also somewhat dependent on 
profile shape \citep{ber01}. 

The real shape of a galaxy is related to its measured shape by dividing its 
ellipticity components $e_i$ by a dilution correction (1 - S$_m$). When 
S$_m$ is large,this correction becomes large and ill determined. As a result, 
we limit our selection of source galaxies to those objects for which
S$_m$ is less than 0.8. This is also an effective star-galaxy
separation technique.

In addition to the isotropic part of the PSF, which dilutes the observed shape
of an object, anisotropic components of the PSF induce ellipticity
in an object's shape. The induced ellipticity is directly proportional to the
ellipticity of the PSF. Because this effect is a convolution of the true 
image with 
the PSF, it will be strong for objects comparable in size to the PSF 
and weaker for larger objects. Again, it is approximately the ratio 
of PSF size to object size that is the important factor. 
We correct for
PSF anisotropy by subtracting off the PSF shape weighted by the 
smear polarizability. The corrected shape of each galaxy is then given
by:
\begin{equation}
e_{i}(corrected) = \frac{e_{i}(observed) - S_{m}*e_{i}(PSF)}{1 - S_{m}}
\end{equation}

SDSS observations are obtained continuously, in drift scan mode. During the
night the PSF varies for many reasons, including changes in
atmospheric seeing, focus, and anomalous refraction. As a result,
the PSF must be carefully tracked throughout the data so that the
appropriate form can be used to correct each galaxy. 
This is a problem which is aided by the relatively shallow nature of the SDSS. 
The numbers of observed stars and galaxies here are similar. In very deep
lensing studies galaxies greatly outnumber PSF stars. 
PHOTO tracks the PSF as a function of position and time using the large 
number of stars in 
each image, allowing accurate PSF reconstruction at the position of 
each galaxy \citep{lup01}.

\subsection{Shear Measurements} \label{shearmeasure}

Lensing induces
tangential shear in the images of background source galaxies. 
As long as the orientation of the distant source galaxies is
otherwise independent of the foreground lens, one can measure
the induced shear directly from the shapes of source galaxies.
For the special case of weak lensing $(\kappa \ll 1)$, this induced
shear is just half the induced tangential ellipticity \citep{Escude91}.
We denote this tangential component of the ellipticity as 
$e_+$ and the orthogonal, 45 degree component of the polarization
as $e_{\times}$.

Because galaxies have intrinsic shapes, and the lensing effect
we are measuring is weak, the shape of a single galaxy
gives a very noisy estimate of the shear. In order to increase
the signal to noise, we average the
tangential shapes of source galaxies in bins of radial 
separation from the foreground lens:

\begin{equation}
\hat{\gamma}_+ = 
\frac{1}{2S_{Sh}} \frac{\sum{w_i e^i_+}}{\sum{w_i}},
\end{equation}
where the $w_i$ are the weights for each measurement. S$_{sh}$ is the
average responsivity of the source galaxies to induced shear (see
below).

There are two dominant sources of random noise in shear measurements: 
the intrinsic variance in galaxy shapes, or ``shape noise'', and the uncertainty 
in the shape measurement of each galaxy. The mean variance in galaxy
ellipticity, determined from well measured SDSS galaxies is
$\sigma_{SN} = \langle e_i^2 \rangle \sim 0.32$, and the typical
shape measurement uncertainty is $\sigma_i \sim 0.25$. 
For optimal S/N we weight the contribution of each source
galaxy by $w_i = 1/(\sigma_i^2. + \sigma_{SN}^2)$.

The shear responsivity measures the efficacy with which an
applied lens shear can alter the observed shape of a galaxy, and is simular
to the shear polarizability defined in \cite{kai95}. If a galaxy
is perfectly tangentially aligned, and has an ellipticity of 1, it
cannot be made to appear more tangentially aligned by the lens. We measure 
the responsivity for the ensemble of sources using the method of 
\citet{ber01}:
\begin{equation}
S_{sh} = \frac{\sum_{i}[w_{i}(1 - \sigma^{2}_{SN}w_{i}e^{2}_+)]}
	{\sum_{i} w_{i}}
\label{combining_lenses}
\end{equation}
Combining this measure of shear responsivity with the weights described 
above yields a measurement of mean shear as a function of angle from the
lens center.

\subsection{Accounting for lens geometry} \label{geometry}

To convert this measurement of shear as a function of angle to a measurement
of mass contrast as a function of physical radius we have to account for the 
lens geometries. The conversion from angular separation to projected physical
separation requires only application of the angular diameter redshift relation.

To place our measurements on an accurate mass scale we must determine
a mean $\Sigma_{crit}$ for each lens.
We do this by combining the measured lens redshifts with
the overall redshift distribution of the background galaxies. 
\begin{equation} \label{meanscrit}
\Sigma_{crit}^{-1}(z_L) = { {\int_{z_L}^{\infty} 
	\Sigma_{crit}^{-1}(z_L,z_s) P(z_s) dz_s} }
\end{equation}
Where P($z_s$) is the normalized source redshift distribution.  

To estimate the source redshift distribution we gather a sample of galaxies
with spectroscopic redshifts and accurate photometry. Two samples are combined.
For galaxies brighter than \rs = 17.6, we use our own SDSS lens galaxy sample.
For galaxies fainter than \rs =17.6, we use galaxies drawn from the 
Canada-France Redshift Survey (CFRS: \cite{Lil95}).  CFRS galaxies have magnitudes measured 
in V and I.  We convert these to \rs\ using the relation
\begin{equation}
r^{\ast} = I_{CFRS} + 0.5(V_{CFRS} - I_{CFRS})
\end{equation}
This relation is empirically determined from 61 galaxies observed by both 
SDSS and CFRS. This direct comparison automatically accounts for average
differences between the isophotal photometry of CFRS and the Petrosian 
photometry of the SDSS. The combined sample gives us a list of calibration 
galaxies with spectroscopic redshifts spanning a range from \rs = 14 to
\rs = 23.

\begin{figure}[t]
\plotone{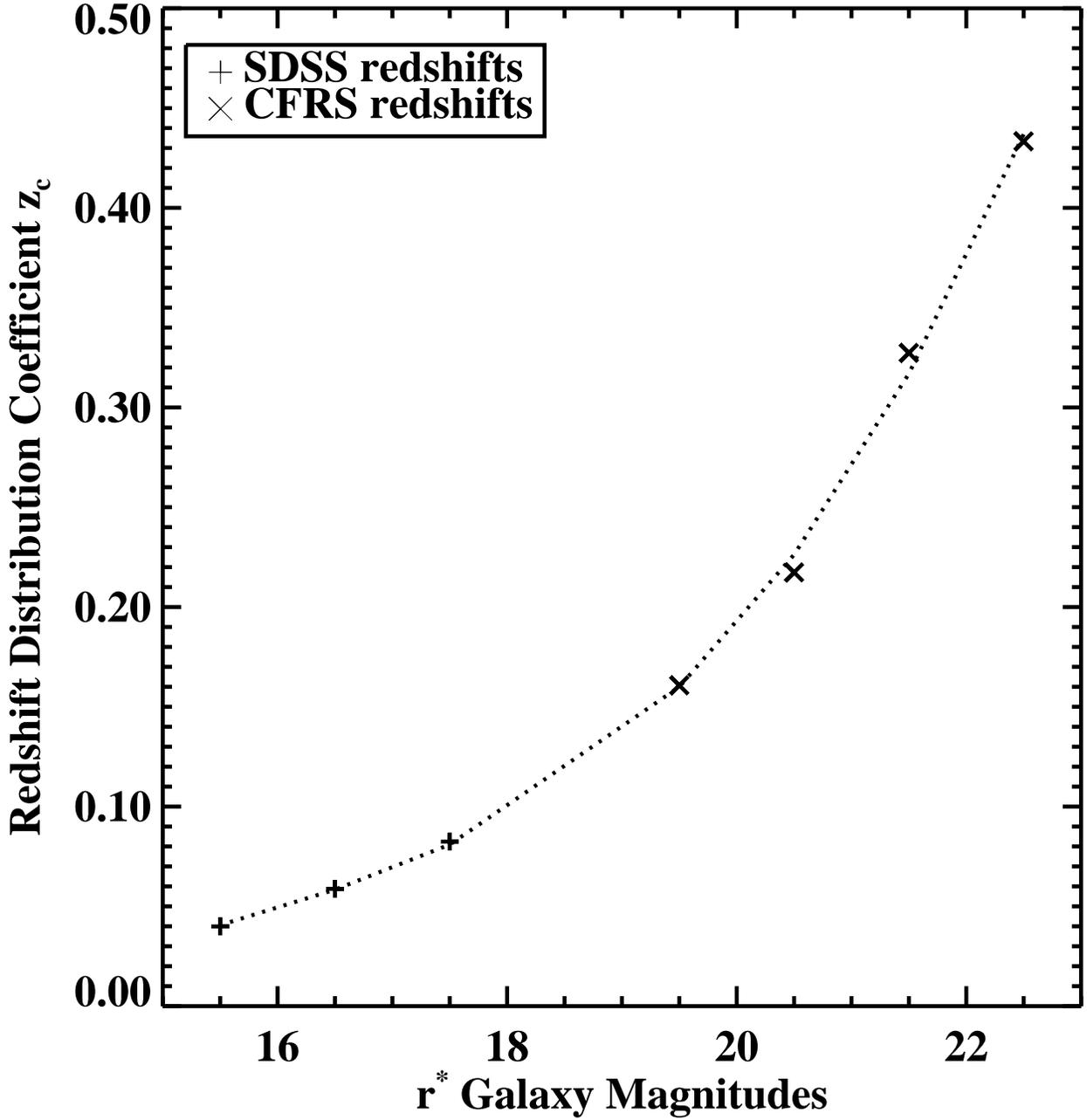}
\figcaption[f4.eps]{This figure shows 
the relationship between the parameter $z_{c}$,
which characterizes the redshift distribution of galaxies, and \rs\
magnitude. Points marked +
represent magnitude bins for which $z_{c}$ was determined
from SDSS galaxies. Points marked x represent magnitude bins for which 
$z_{c}$ was determined from CFRS galaxies. The dashed line shows the 
best fit $z_{c}$ vs. magnitude relation used in determining the source 
redshift distribution.
\label{zc_vs_mag}}
\end{figure}

\begin{figure}[t]
\plotone{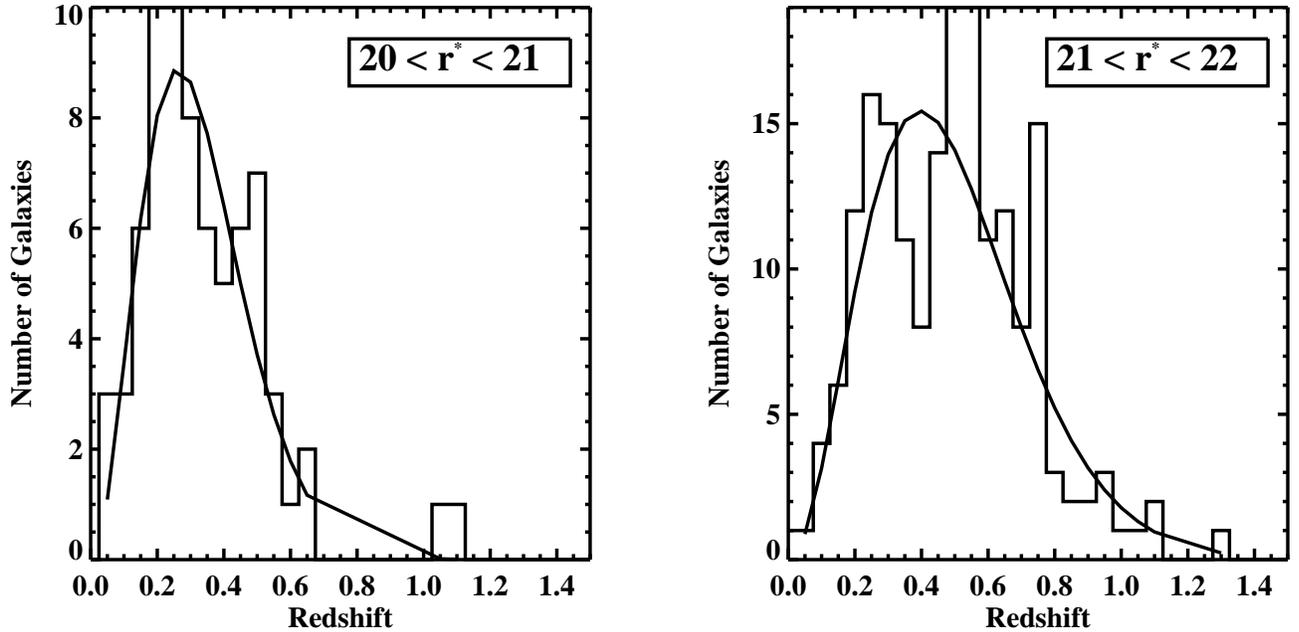}
\figcaption[f5.eps]{The accuracy of source redshift 
estimation is suggested by this comparison between the real CFRS spectroscopic
redshift distribution (histogram) and the redshift distribution estimated
by the method described in \S \ref{geometry} (solid line). \
The comparison is made for 
galaxies with 20 $<$ \rs\ $<$ 21 on the left, and for galaxies with
21 $<$ \rs\ $<$ 22 on the right.
\label{source_estimate_vs_z}}
\end{figure}

To estimate the source redshift distribution as a function of magnitude 
we fit the redshift distributions of the calibration galaxies, in bins of 
\rs, to the function (\cite{Baugh93})
\begin{equation} \label{zfit}
n(z) = \frac{3z^2}{2z_{c}^3} \textrm{exp}(-\left(\frac{z}{zc}\right)^{3/2})
\end{equation}
The fit values of $z_{c}$ in each \rs\ bin are then used to derive a relationship 
between \rs\ and
$z_{c}$. We combine this relation with the observed \rs\ magnitude 
distribution of
the source galaxies to obtain a source galaxy redshift distribution. 
The relation between \rs\ and $z_{c}$ is shown in Figure \ref{zc_vs_mag}.
As an illustration of the effectiveness of this method we compare real and
estimated redshift distributions for CFRS galaxies in Figure 
\ref{source_estimate_vs_z}.

It is worth noting that, since the CFRS is complete to m$_I$ of 22.5, no 
extrapolation of the \rs\/$z_{c}$ relation
to magnitudes fainter than the calibration are
required. Lens geometries are, on average, understood very well for this 
analysis. In addition, since the mean number of source galaxies per lens is 
$\gtrsim$100, each lens samples this distribution relatively well.

\begin{figure}[t]
\plotone{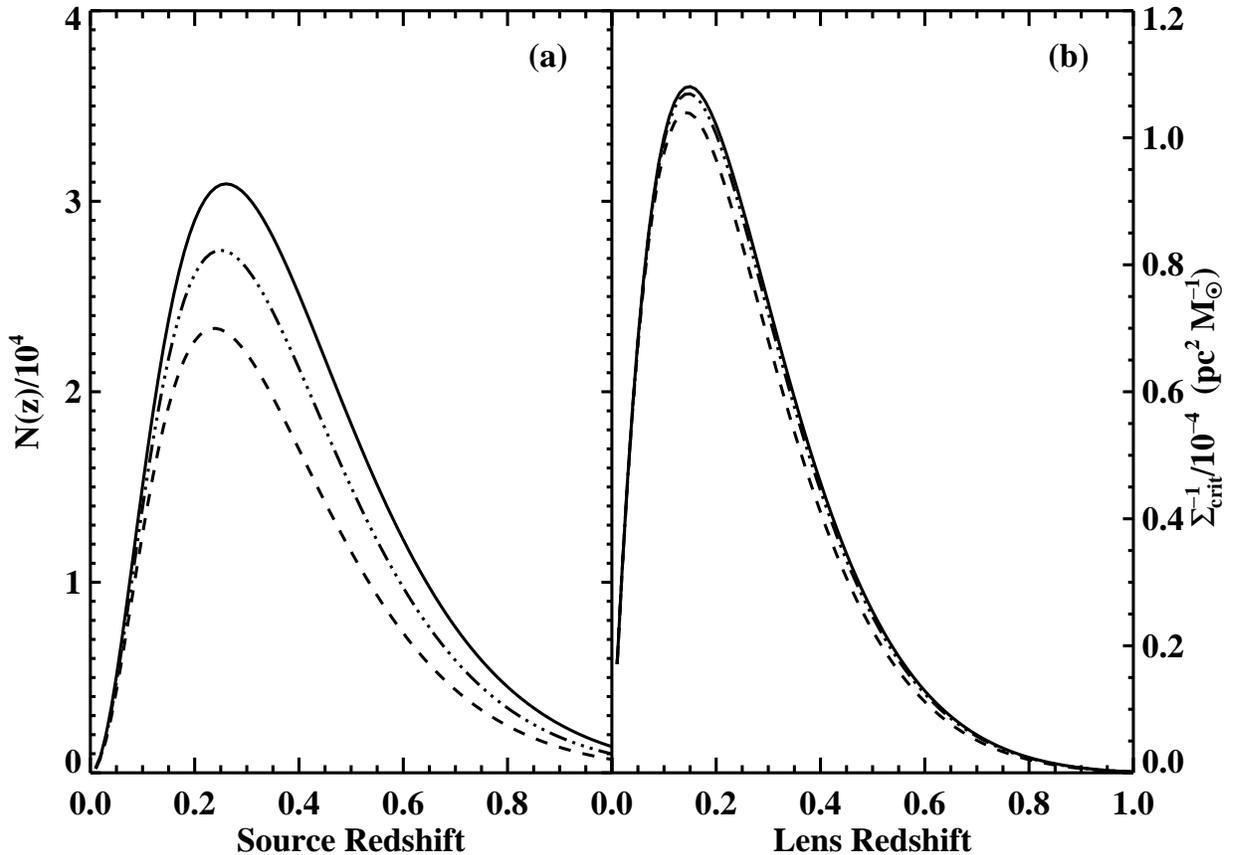}
\figcaption[f6.eps]{Estimated source galaxy
redshift distribution in bins of .01 (a) 
and inferred lensing strength $\Sigma_{crit}^{-1}$
as a function of lens redshift (b). Source galaxies are taken from the 
\gp (dashed line), \rp (solid), and \ip (dot-dashed) images.
\label{nzandscrit}}
\end{figure}

The inferred source redshift distribution is shown in
figure \ref{nzandscrit}(a). The effect of geometry on a lens can be expressed
as a ``lensing strength'' characterized by $\Sigma_{crit}^{-1}$. Where 
$\Sigma_{crit}$ is small, so that $\Sigma_{crit}^{-1}$ is large, lensing
effects will be large. The mean lensing strength as a function of
lens redshift, shown in figure \ref{nzandscrit}(b), is calculated from the 
source redshift distribution using equation \ref{meanscrit}. Distributions of
both are shown for \gp, \rp, and \ip\ because the source samples in each
band are slightly different.

\subsection{Combining lenses}\label{combining}

The sensitivity of shear measurements goes as $\sigma/\sqrt{N}_{source}$
where $\sigma = \sqrt{\sigma_{sh}^2 + \sigma_{meas}^2} \sim 0.4$.
The maximum ellipticity induced by a SDSS
foreground galaxy on background sources 
is very small, $\lesssim$ .005 \citep{fis00}. At least 10 thousand 
source galaxies are required to measure such a small signal.  
SDSS images, however, typically have $\sim$ 2 sources per
square arcminute, or about 650 within a 600$^{\prime\prime}$ aperture. 
This is too few sources to measure the shear from a single 
galaxy with precision. In order
to obtain the required number of sources, we combine the signal from many
lenses.

Care must be taken when combining the shear measurement from lenses
at different redshifts. As equation \ref{tshear} suggests, the shear 
depends upon the redshift of the lens and source. If we wish to combine 
lenses, we must combine measurements of density contrast (Equation 
\ref{densconteq}) rather than measurements of shear (Equation 
\ref{tshear}). Note that the weights used in equation 
\ref{combining_lenses} must also be rescaled to 
$W_i$ = $w_i\Sigma_{crit}^{-2}$.

On average, equation \ref{meanscrit} accounts for the geometry of the
lens-source system, including the fact that some sources are actually
in front of the lens and are not lensed. This average relation does not,
however, account for the intrinsic clustering of galaxies.
Some of the sources are faint neighbors physically associated with 
the lens galaxies. These faint neighbors are not lensed. 
Because their number density is a decreasing function of distance from 
the lens center, this will produce a radial bias in the shear estimate.
A comparison of the number density of sources around lenses to the number
density around random points gives an estimate of the mean number of faint
galaxies associated with the lenses. 

This dilution effect can be corrected for
by multiplying the mean density contrast by a radial correction factor
F(r) = $\langle N_{lens}\rangle/\langle N_{rand}\rangle$, the ratio
of source galaxies around lenses to that around random points.
We chose 155,000 random points (5 random points for each lens), 
spread uniformly through out the same 
areas sampled by the lenses and drawn from the same redshift distribution 
as the lenses, in order to measure the mean background density correctly.
The correction F(r) varies from 6\% at 70 \hinv kpc 
to 0.04\% at 940 \hinv kpc.

\section{The Galaxy-Mass Correlation Function}\label{galmasscorr}

\begin{figure}[t]
\epsscale{0.8}
\plotone{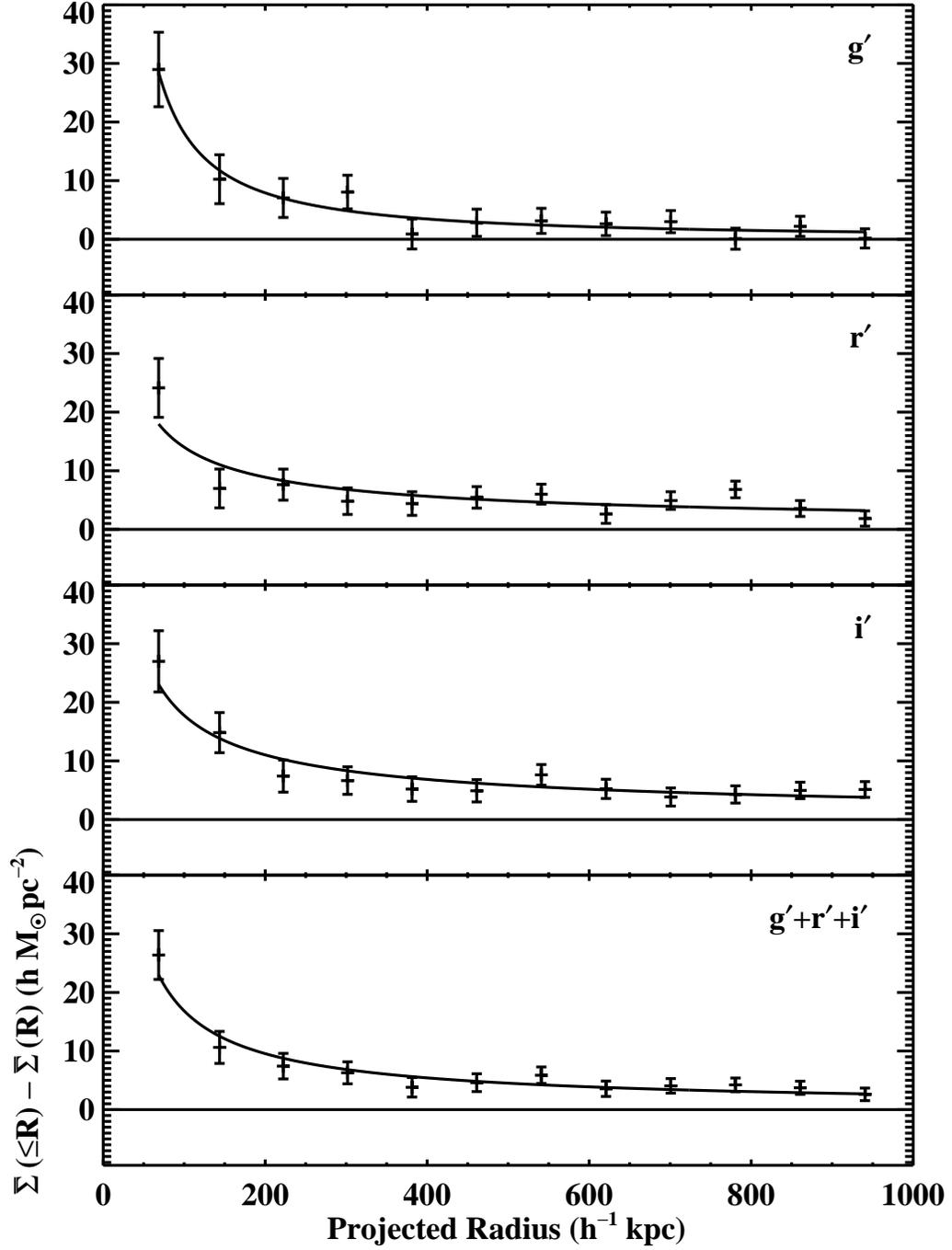}
\figcaption[f7.eps]{Mean density contrast
measured as a function of projected radius around 
$\sim$31,000 SDSS lens galaxies. The plots are the mean density contrast
in \gp, \rp, and \ip\ images from the top, with the combined data on bottom.  
The solid lines are the best-fitting power laws.
\label{all_lens_profile}}
\end{figure}

The ensemble density contrast measured around the
lens galaxies is shown in Figure \ref{all_lens_profile}. Measurements
based on source galaxy shapes in each of the three most sensitive bandpasses 
are shown. These profiles include corrections for the shear responsivity as 
well as the correction for the inclusion of faint neighbors in the 
source galaxy sample discussed above. Only lenses which were surrounded by a
reasonably symmetric distribution of source galaxies were used. This check
is done to allow residual correlations in source galaxy shapes to cancel
as they are projected tangentially around the lens centers.
The final sample includes 31,039, 31,174, and 31,203 lens galaxies for 
the \gp, \rp, and \ip\ measurements, respectively.

Figure \ref{all_lens_profile} also shows a the measurement of the GMCF 
obtained by combining the measurements made in \gp, \rp, and \ip.
Lensing measurements made in these three bands are not entirely independent.
First, they measure shapes of the same source galaxies. While the intrinsic
shapes of these galaxies are not exactly the same in these different bands,
they are strongly correlated. So the shape noise in these three measurements
is related. While each color has a PSF shape independent of the others,
the time variation of the PSF (which is largely a global, seeing effect) is
correlated among the bands. As a result, the combination of the three bands
requires consideration of the correlation matrix among the colors, and
yields gains in S/N which are less than the optimal $\sqrt{3}$.

\begin{figure}[t]
\plotone{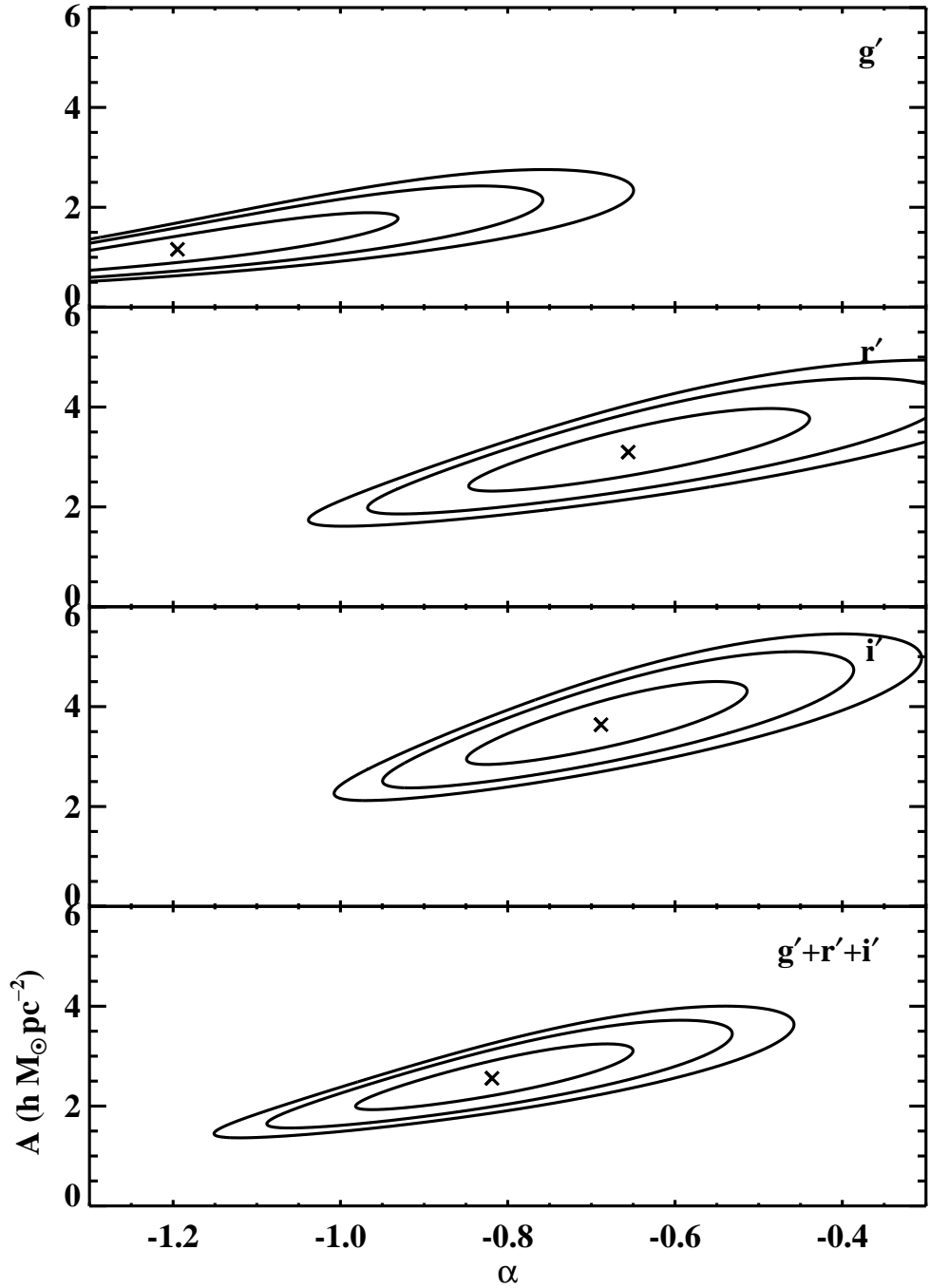}
\figcaption[f8.eps]{Best fits of the GMCF to the function 
$\Delta\Sigma_{+} = A (R/1\ \textrm{Mpc})^{-\alpha}$ are shown for each 
band and for the combined data. $\chi^{2}$ contours for each represent
68\%, 95\% and 99\% confidence.
\label{all_lens_fits}}
\end{figure}

The density contrast measurements displayed in Figure \ref{all_lens_profile}
are fit to a power law model of the form:
\begin{equation}
\Delta\Sigma_{+} = A \left(\frac{R}{1\ \textrm{Mpc}}\right)^{-\alpha}
\end{equation}
Best fit values for the normalization A and the power law index $\alpha$
are given for each color, and the combined measurement, in Table 
\ref{gmcftable}. The $\chi^{2}$ statistic for the combined data 
was estimated using the full covariance matrix; $\chi^{2}$ contours for
these best fits are shown in Figure \ref{all_lens_fits}. 
The best fit to the combined GMCF is 
\begin{equation}
\Delta\Sigma_{+} = (2.5^{+0.7}_{-0.6}) 
	\left(\frac{R}{1\ \textrm{Mpc}}\right)^{-0.8\pm0.2}h 
         \textrm{M}_{\sun} \textrm{pc}^{-2}
\end{equation}
When we study the variation of the GMCF with luminosity in \S 
\ref{gmcf_luminosity} we will fix the power law index $\alpha$ to
this best fit value of 0.8 and allow the normalization A to vary.
All subsequent measurements of the GMCF discussed in this work are 
made from the combined \gp, \rp, and \ip\ data for
maximum S/N.

This measurement of the mean density contrast averages over galaxies 
of many different types, drawn from many different environments. 
In this sense, the mean density contrast is really a measurement of the 
projected galaxy-mass correlation function (GMCF). Included in this 
profile is not only the lensing induced by
the central galaxy, but also the effect of all the neighboring lenses 
either in projection or physically associated with the central galaxy.

\subsection{Checks of the GMCF} \label{gmcf_systematics}

\begin{figure}[t]
\plotone{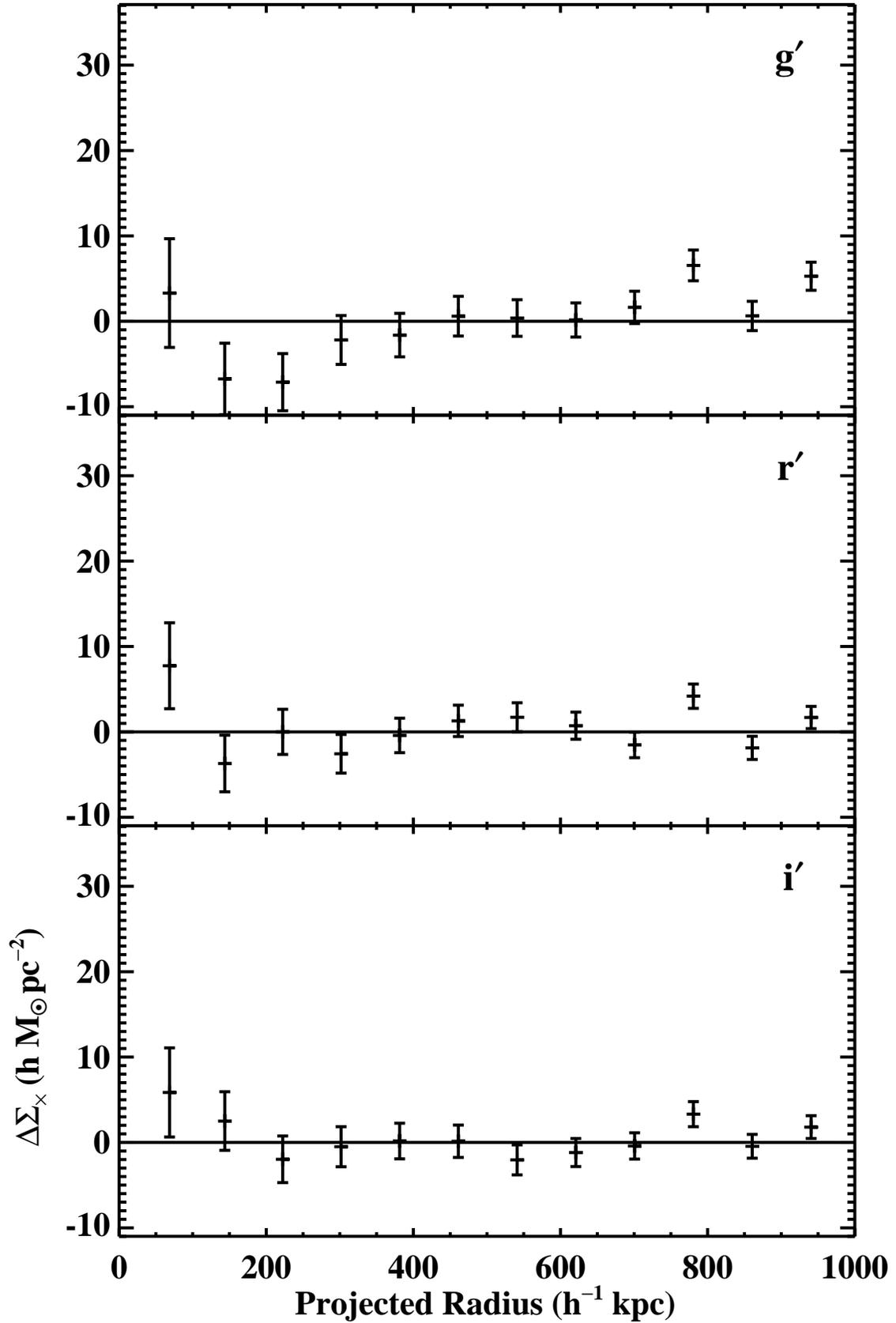}
\figcaption[f9.eps]{
The GMCF is measured here around the same SDSS lenses as in
Figure \ref{all_lens_profile},
but source galaxies have been rotated by 45$^{\circ}$. This rotation
eliminates the lensing signal observed in Figure \ref{all_lens_profile}.
\label{orthodenscont}
}
\end{figure}

As a simple test to see if our signal is due to gravitational lensing
we rotate the orientation of all source galaxies by 45 degrees and
measure the orthogonal of the density contrast, which we
will denote as $\Delta \Sigma_{\times}$. This is
equivalent to measuring the curl of a gradient and should be zero
\citep{kai95a} for any real shear signal. This measurement is shown in 
Figure \ref{orthodenscont} in the same radial bins as the measurement of 
$\Delta \Sigma_+$.

\begin{figure}[t]
\plotone{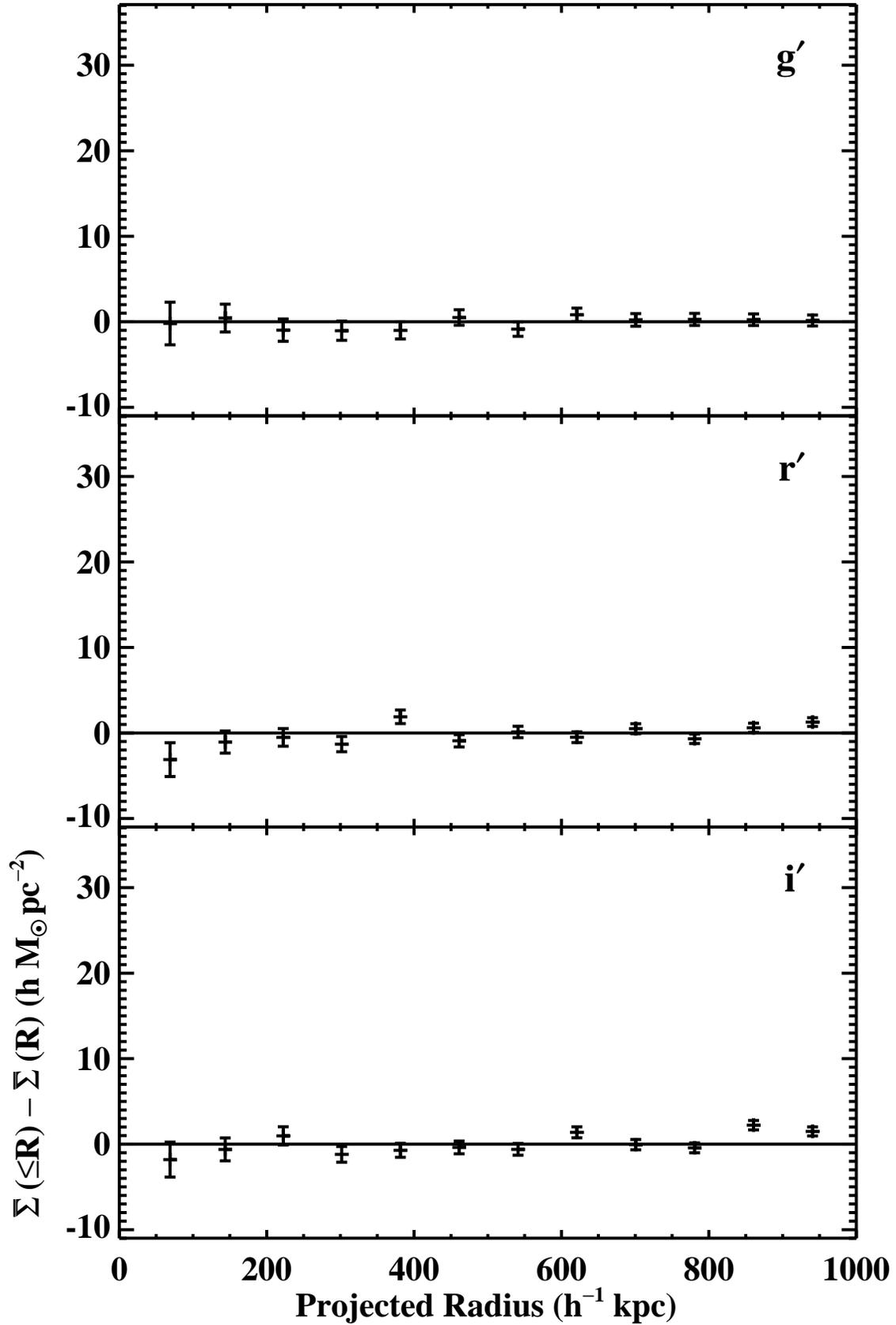}
\figcaption[f10.eps]{
The GMCF is measured here around 150,000 random points.
Consistency with zero surface mass density contrast around these points
confirms that the signal observed in Figure \ref{all_lens_profile} is
associated with galaxies.
\label{denscontrand}
}
\end{figure}

A very powerful check of the level of our systematics can be done by measuring 
the shear around random points. This test is possible because the SDSS 
observes
large contiguous regions of sky. There should be no correlation between 
source galaxy 
shapes and these random points, and the resulting tangential shear should be
consistent with zero. We use the 155,000 random points discussed in
\S \ref{combining} to make this measurement. The result, shown in 
Figure \ref{denscontrand}, is consistent with zero, and far below the measured
signal.

These results demonstrate that the shape correlation which appears in our
measurement of the galaxy-mass correlation function (Figure 
\ref{all_lens_profile})
is consistent with lensing. There is no net tangential alignment of background 
galaxies detected unless you measure its correlation with the location 
of foreground lenses.

If the formation of luminous galaxies is at all affected by the dark matter 
environment in which they form, there should be relationships between the
luminous properties of galaxies and their GMCF. The following sections
present measurements of the dependence of the GMCF on galaxy luminosity,
morphology, and local density. They show strong changes in the amplitude
and in some cases the shape of the GMCF as the luminous properties of the 
lens sample are varied.

\subsection{Dependence of the GMCF on lens galaxy luminosity}
	\label{gmcf_luminosity}

We begin by probing the scaling between galaxy luminosity and the GMCF.
Since the SDSS provides five color photometry, and because we have 
spectroscopic redshifts for every lens galaxy, we can divide the lens
sample by luminosity in each of the five SDSS bands. For each band, we
split the lens sample into four luminosity bins. 

Because of the shape of the galaxy luminosity function \citep{bla01}, we 
expect a large number of low
luminosity galaxies, and relatively small numbers of high luminosity 
galaxies.
This provides greater lensing sensitivity for low luminosity galaxies. 
Since we expect these to have smaller masses, we will need this higher 
sensitivity. 
We choose a division of the lens galaxies which is uneven in number.
This is appropriate for obtaining comparable signal to noise in each bin.
Luminosity bins are chosen to give a similar division of the total number of 
galaxies in each of the five colors; e.g. the third luminosity bin
has approximately the same number of galaxies in the \us\ binning as 
the \rs\ binning. Mean luminosities in each bin are calculated using the 
same relative weighting used in the lensing measurements 
($N_{sources}*\Sigma_{crit}^{-2}$).

\begin{figure}[t]
\plotone{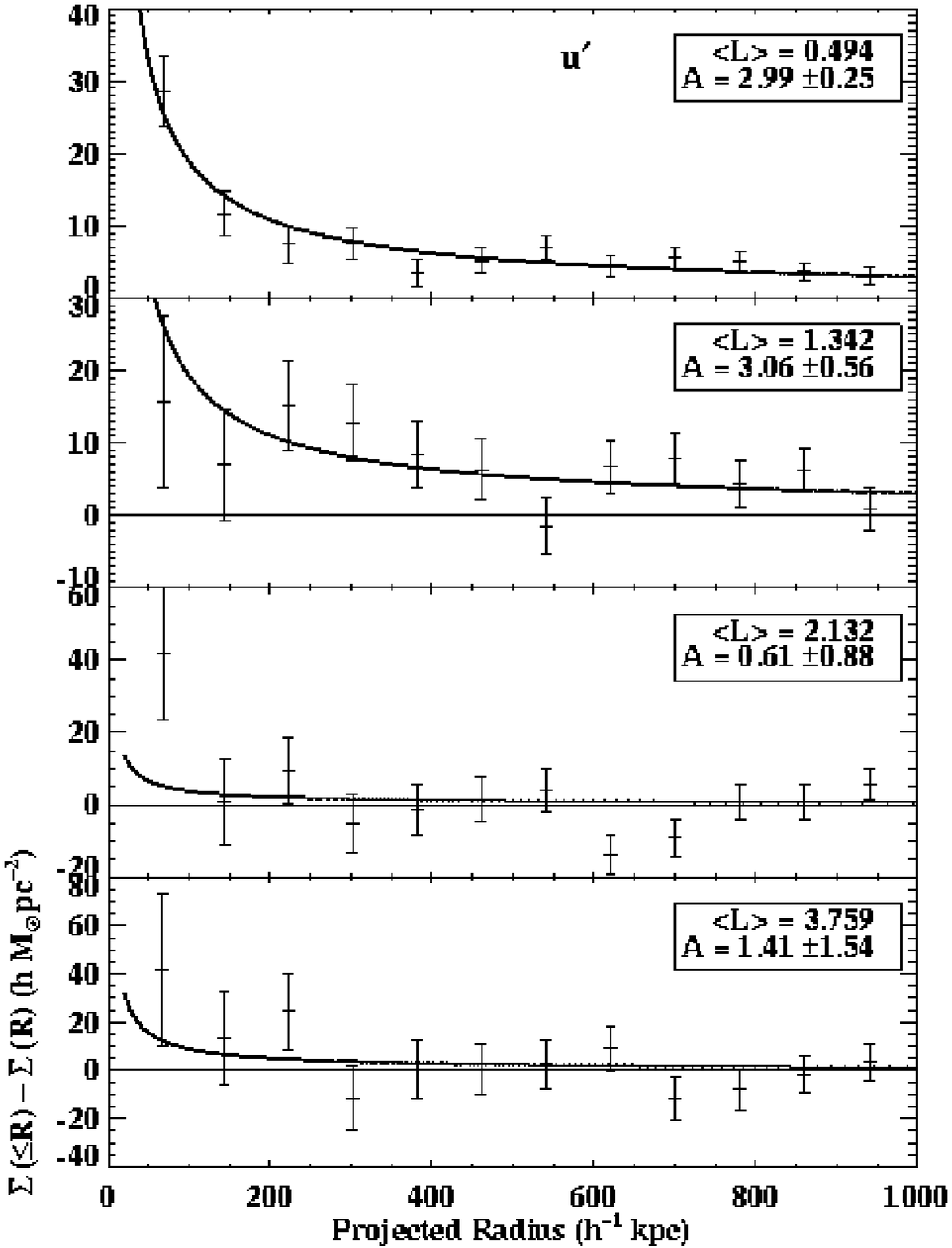}
\figcaption[f11_sm.ps]{
GMCF in bins of \us\ luminosity. Lines are the best fits to 
$R^{-0.8}$ power law models corresponding to the noted values of the 
normalization A. Note that each plot has a different y-axis scale.
\label{denscont_u}
}
\end{figure}

\begin{figure}[t]
\plotone{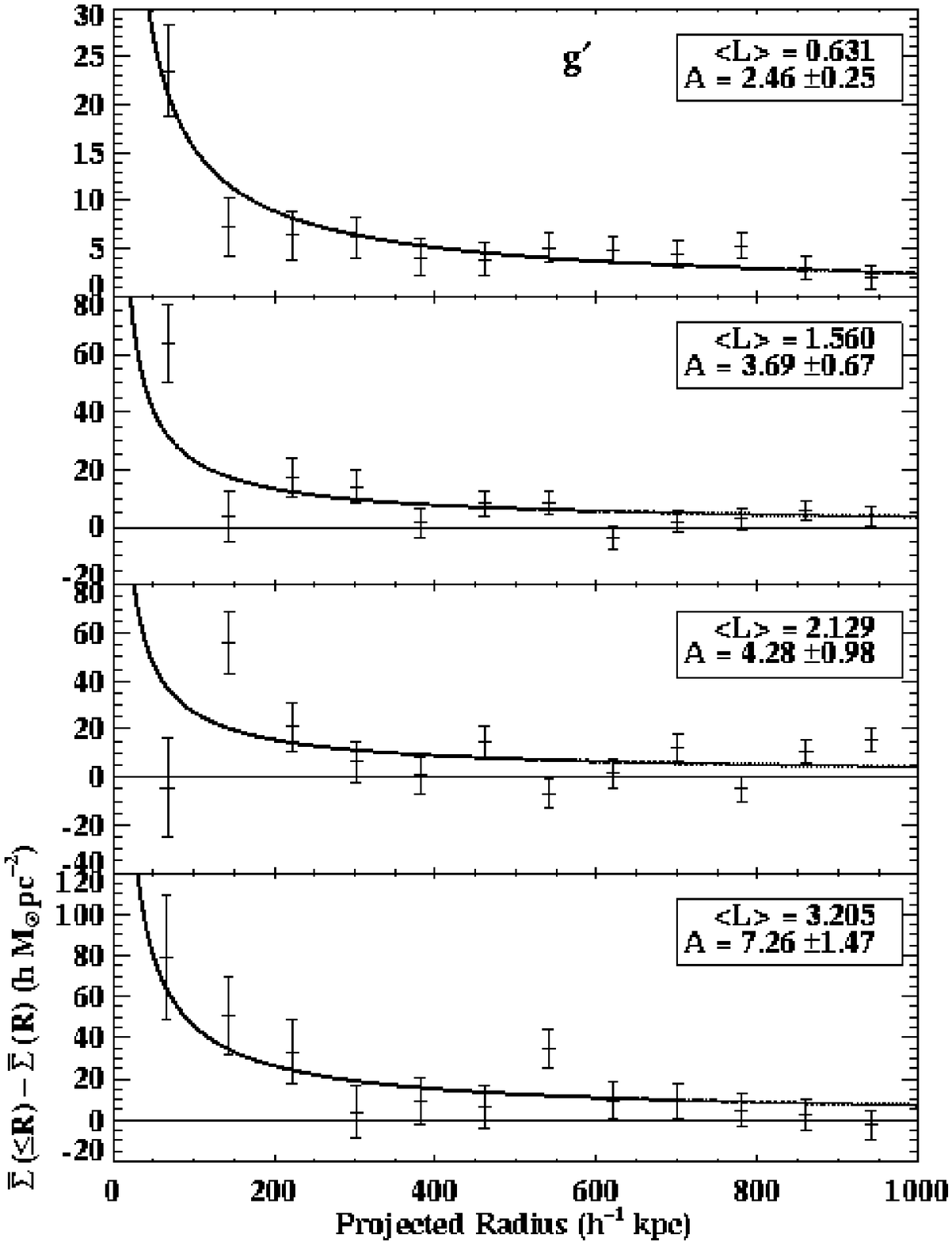}
\figcaption[f12_sm.ps]{
GMCF in bins of \gs\ luminosity. Lines are the best fits to 
$R^{-0.8}$ power law models corresponding to the noted values of the 
normalization A. Note that each plot has a different y-axis scale.
\label{denscont_g}}
\end{figure}

\begin{figure}[t]
\plotone{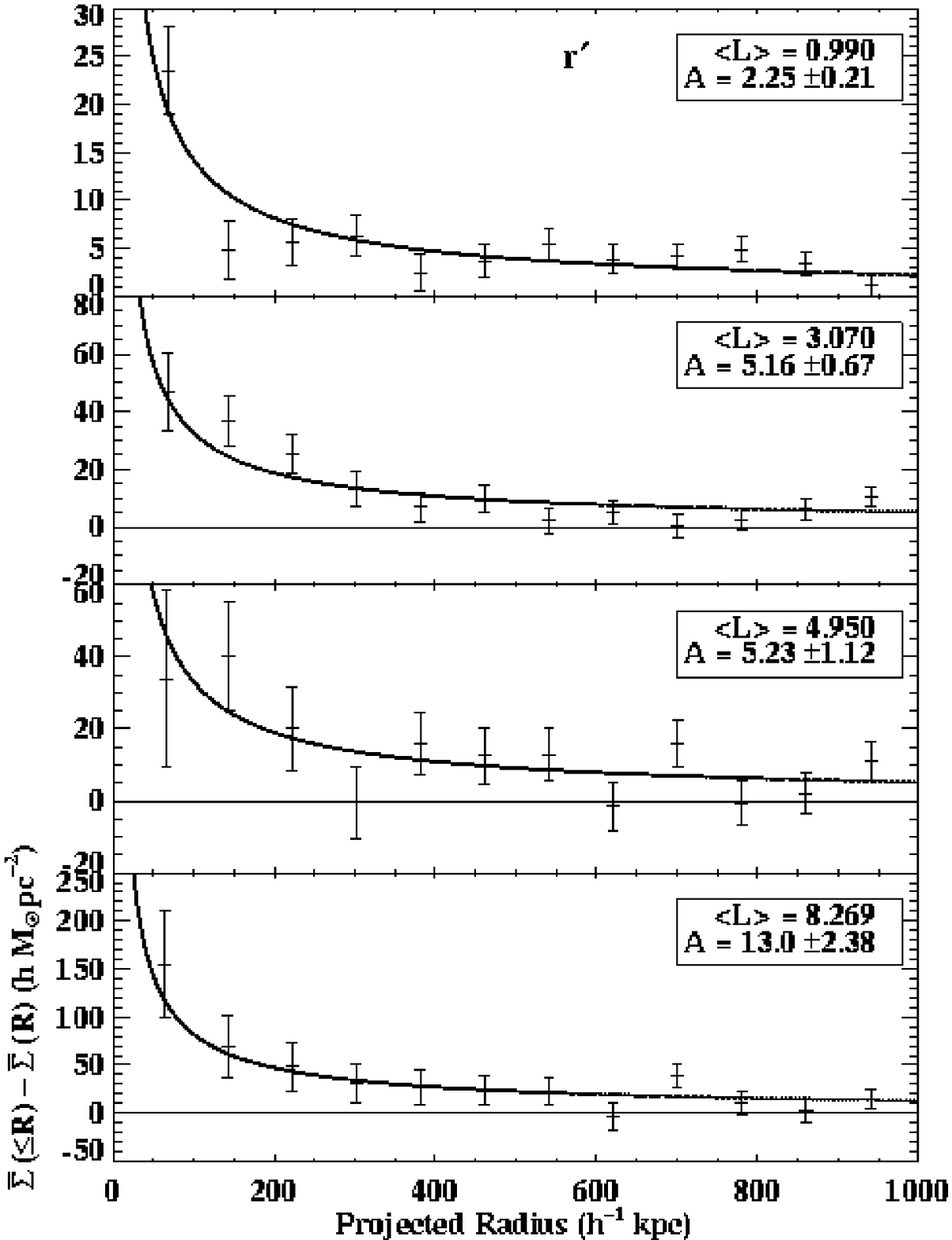}
\figcaption[f13_sm.ps]{
GMCF in bins of \rs\ luminosity. Lines are the best fits to 
$R^{-0.8}$ power law models corresponding to the noted values of the 
normalization A. Note that each plot has a different y-axis scale.
\label{denscont_r}}
\end{figure}

\begin{figure}[t]
\plotone{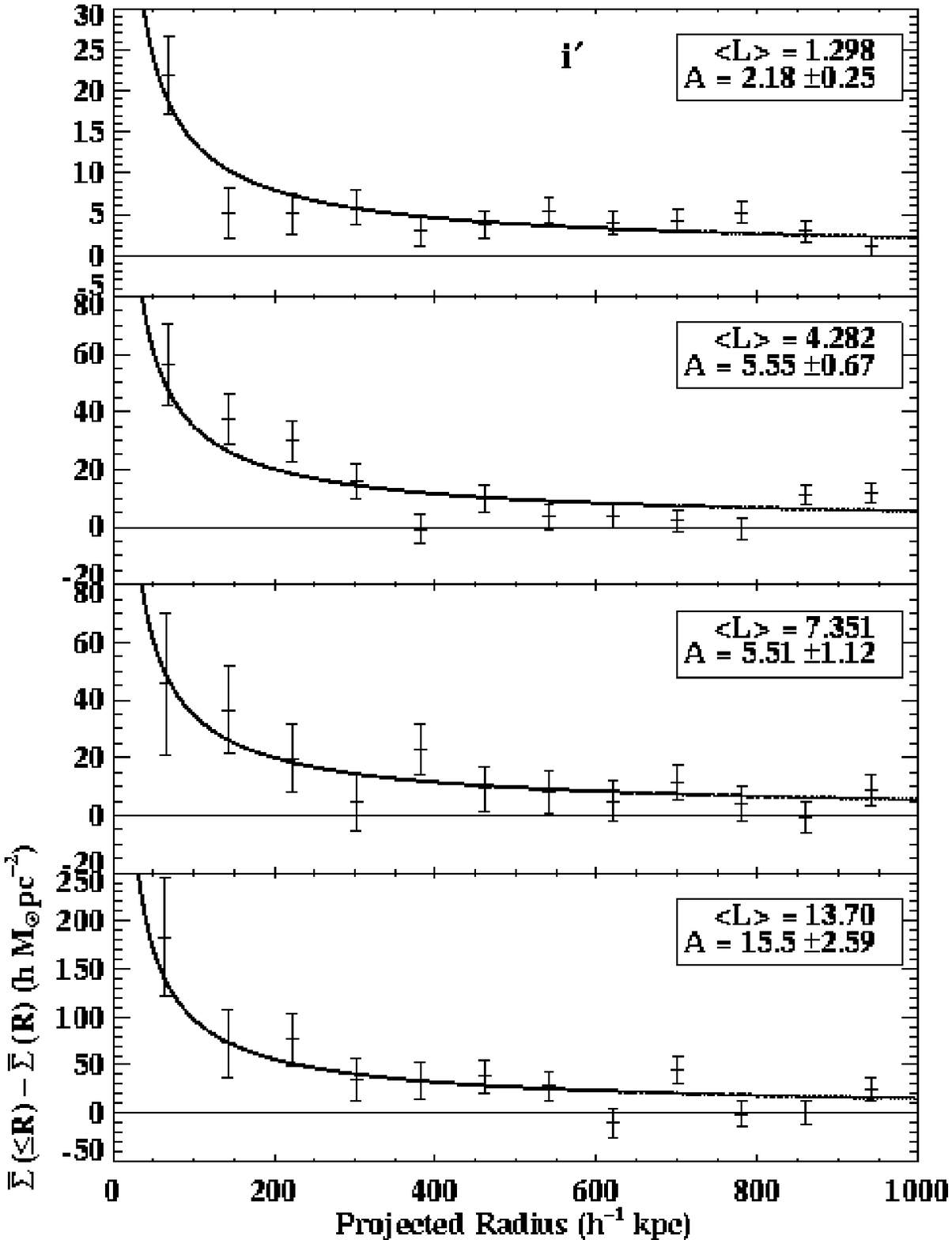}
\figcaption[f14_sm.ps]{
GMCF in bins of \is\ luminosity. Lines are the best fits to 
$R^{-0.8}$ power law models corresponding to the noted values of the 
normalization A. Note that each plot has a different y-axis scale.
\label{denscont_i}}
\end{figure}

\begin{figure}[t]
\plotone{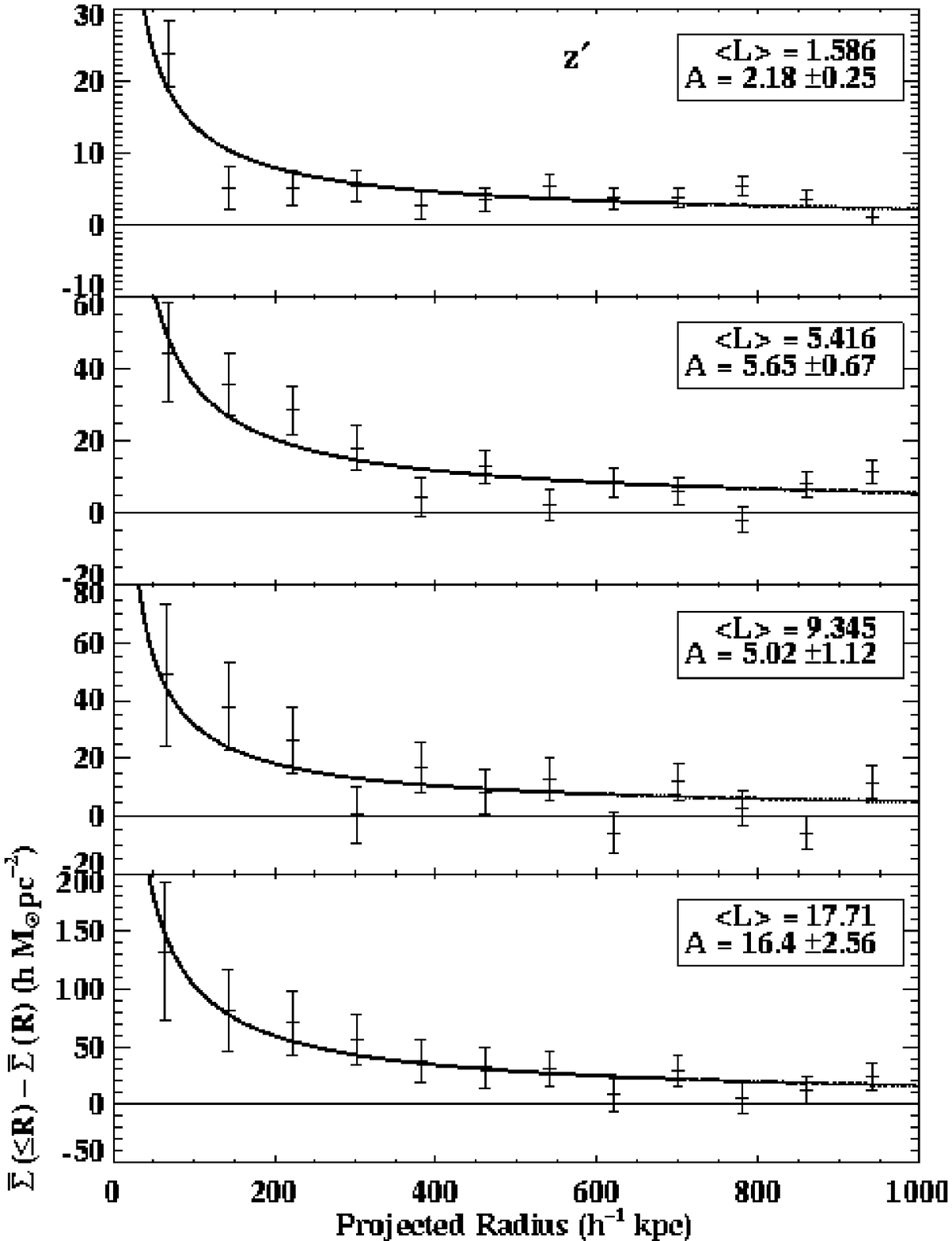}
\figcaption[f15_sm.ps]{
GMCF in bins of \zs\ luminosity. Lines are the best fits to 
$R^{-0.8}$ power law models corresponding to the noted values of the 
normalization A. Note that each plot has a different y-axis scale.
\label{denscont_z}
}
\end{figure}

Figures \ref{denscont_u} through \ref{denscont_z} show the GMCF split into
these four luminosity bins, using \us, \gs, \rs, \is, and \zs\ luminosities
respectively. The vertical scale of these plots varies to accomodate the 
large change in the amplitude of the GMCF between bins, especially in the 
redder bands. In each case, we present the GMCF measured from a combination
of all three source passbands. The GMCF in each luminosity bin is fit to
the model GMCF obtained for the total data set:

The plots are labelled with their mean
luminosity in that band, as well as a normalization of the GMCF from fits
to a power law with index -0.8, the best-fitting power law index for the
entire sample.

There is a strong increase in the amplitude of the GMCF with
\gs, \rs, \is, and \zs\ luminosity, but there is little correlation between
the GMCF and the \us\ luminosity. Put another way, the luminosity of galaxies 
measured in redder bands is strongly related to the dark matter environment 
in which they form, while the \us\ luminosity is not. This is probably not 
surprising. The \us\ luminosity is primarily due to small scale, 
relatively brief episodes of recent star formation, while the \rs, \is, 
and \zs\ luminosity arises from the integrated star formation history of 
the galaxy. It appears that this integrated star formation history is 
much more closely related to a galaxy's dark matter environment than 
its recent star formation history.

Table \ref{lumbintable} presents the essential data. For each luminosity 
bin, in each color, we supply
the number of galaxies used, the mean luminosity of the galaxies, and the 
measured density contrast in the annulus 20 $\le$ R $\le$ 260 \hinv kpc. 
While the amplitude of
the GMCF is increasing strongly with luminosity, the number of available 
lenses is decreasing, so that the S/N of our GMCF measurement is very similar
in each of our four luminosity bins (with the exception of \us). Note that 
we observe no significant change in the {\it shape} of the GMCF with 
luminosity, only the amplitude is seen to change.

\subsection{Dependence of the GMCF on lens galaxy type} \label{gmcf_type}

We have also examined the dependence of the GMCF on galaxy type. Galaxies are 
divided into an early type `elliptical' sample and a late type `spiral' sample
as described above in \S \ref{classification}. The sample used includes
13,882 objects classified as spirals and 14,156 classified as ellipticals. 
The relatively large fraction of elliptical galaxies is due to the magnitude
limited SDSS spectroscopic target selection.

\begin{figure}[t]
\plotone{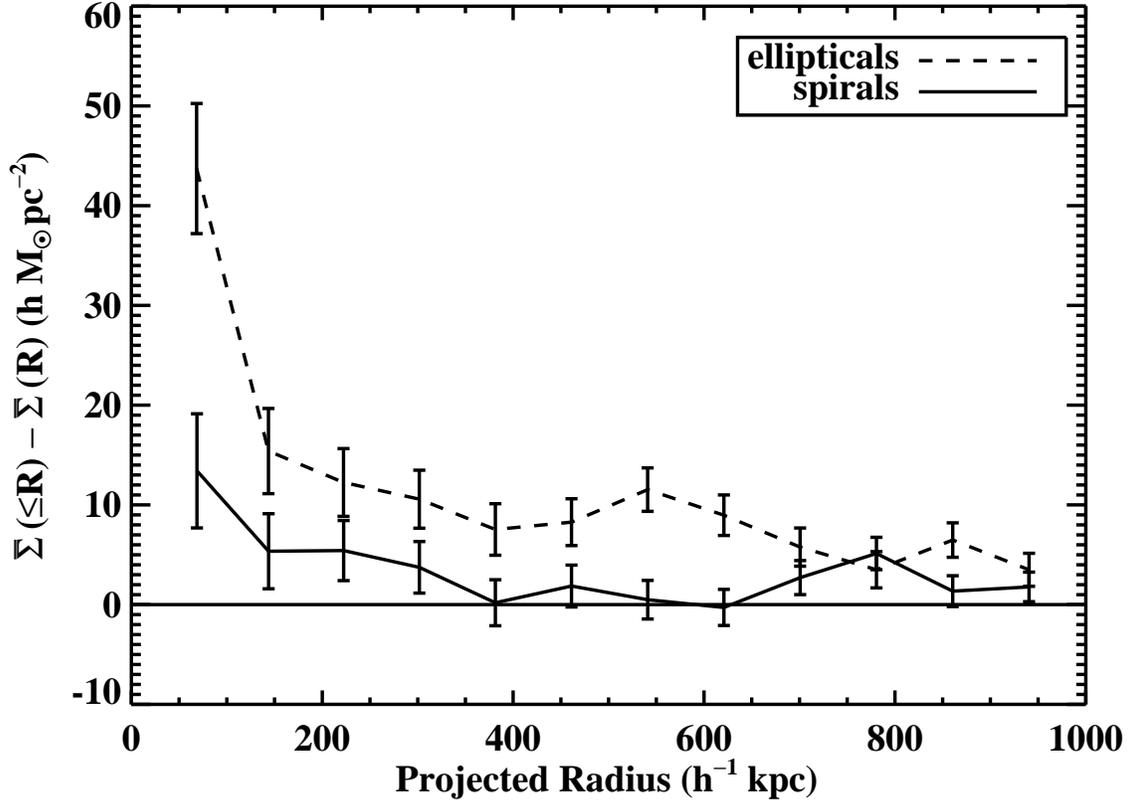}
\figcaption[f16.eps]{
This figure compares the GMCF measured around spiral and elliptical galaxies.
Sample selection is described in \S \ref{classification}. 
\label{denscont_type}
}
\end{figure}

The GMCF measured for each of the two samples
is shown in Figure \ref{denscont_type}. Power law fits to these measured 
GMCFs are given in Table \ref{gmcftable}.
The amplitude of the GMCF for 
elliptical galaxies is substantially larger than that for the spirals.
The shapes of these two GMCFs are consistent, although there
is some evidence for environmental effects at large radii 
(see \S \ref{gmcf_environment}). 
It is important to stress that the GMCF is a redshift independent 
quantity, unlike the induced shear.
Thus the observed discrepancy in central values of the GMCF between spirals 
and ellipticals reflects a real difference in mass.

This measurement indicates that low redshift elliptical galaxies reside 
in halos which are substantially more 
massive than those of spirals. This is perhaps consistent with a picture
in which elliptical galaxies form in the merger of two or more {\it 
comparably massive} objects \citep{bar99,bur00} and so tend to be 
found in more massive halos. Thus elliptical lenses 
will reside in locations with particularly strong GMCFs. While these 
ellipticals are on average more massive than the spirals, they are also
more luminous. Details will be given in \S \ref{m2l}.

\subsection{Dependence of the GMCF on lens galaxy environment}
	\label{gmcf_environment}

In an effort to understand the relationship between GMCF and the environment
of the lens galaxy, we have measured the GMCF as a function of local lens
density. To do this, we generate a Voronoi tesselation \citep{ick87,ling87}
around the locations
of the lens galaxies. The Voronoi tesselation divides the plane of galaxy 
locations into polygons. Each point in the polygon which contains a lens 
is closer to {\it this} lens than to any other. Hence each Voronoi 
polygon reflects the area `owned' by this lens. The inverse of the 
area of each Voronoi polygon is a reflection of its local density. Small
Voronoi polygons occur in the regions of largest local galaxy density.

\begin{figure}[t]
\plotone{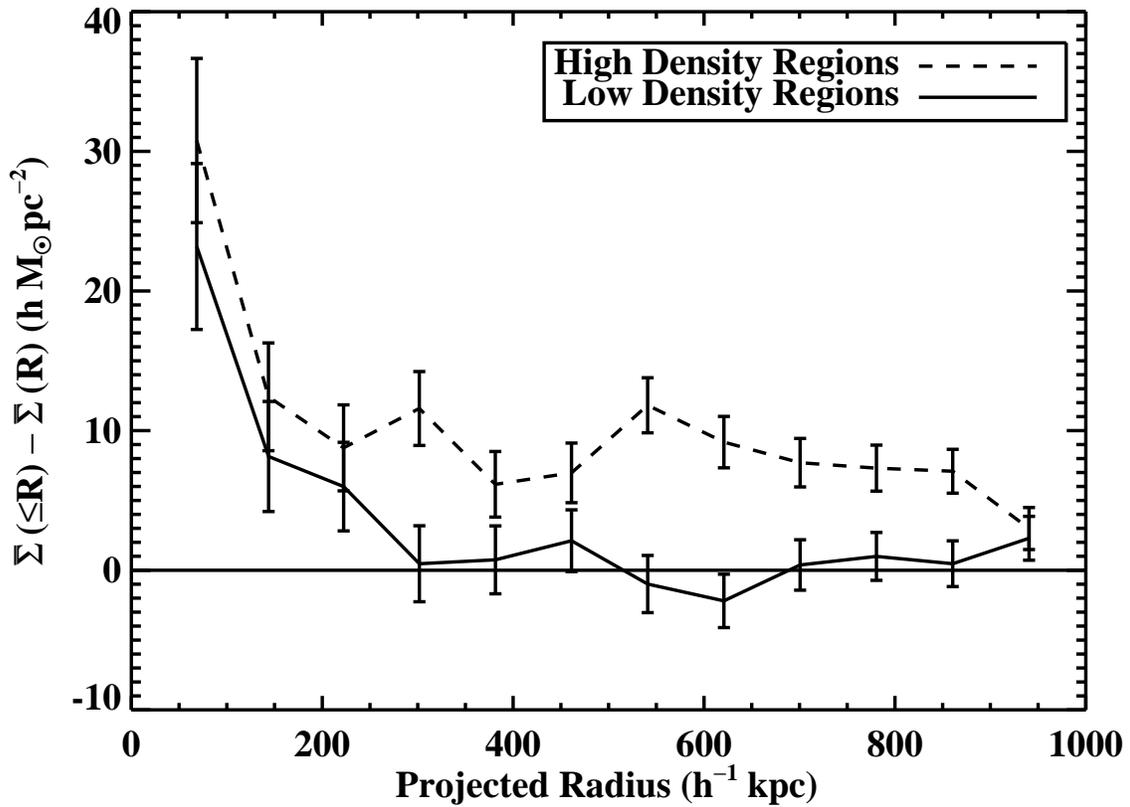}
\figcaption[f17.eps]{
This figure compares the GMCF measured around galaxies in high and low density
regions. Sample selection is described in \S \ref{gmcf_environment}.
\label{denscont_environment}
}
\end{figure}

We divide the lenses into equal samples drawn from overdense and 
underdense regions and measure the GMCF for each. The resulting GMCFs are
shown in Figure \ref{denscont_environment}. While the comparison of spirals 
and ellipticals showed
us a variation in the GMCF amplitude with no change in shape, the comparison
of samples drawn from dense and underdense regions shows a variation in
GMCF {\it shape}, with no significant change in the central amplitude. 
Parameters for best fit power laws to these GMCF measurements are given
in Table \ref{gmcftable}.
The inner parts
of the GMCF for these two samples are statistically equivilant. They differ
only in their outer regions, where the GMCF measured in dense regions is
substantially larger. 

The dense and underdense samples include lens galaxies whose intrinsic 
properties (luminosity and morphological type) are very similar. The mean
\is\ luminosity of the two samples is 2.0 and 2.1 $\times 10^{10} L_{\sun}$
respectively. These lenses
differ significantly only in their surroundings. The similarity of their 
observed GMCFs at small radii suggests that this inner part of the GMCF is 
primarily associated with the central galaxy. The differences observed 
at large radii are due to the presence of many more neighbors around the
galaxies drawn from the dense sample.

These comparisons provide important clues to interpretation of the GMCF. The 
central parts of the GMCF are dominated by mass associated with the central
galaxy. The outer regions (beyond $\sim$300 \hinv kpc) are increasingly 
affected by the presence of neighboring objects, and hence depend on 
the properties of the central galaxy in at most a second order way.

\section{Aperture mass measurements and mass-to-light scalings} \label{m2l}

We have seen in the previous sections that the amplitude of the GMCF is 
strongly dependent on lens luminosity and type, and that the shape of the 
of the GMCF, especially beyond $\sim$300 \hinv kpc is affected by 
the local environment of the lens. The arguments presented in \S 
\ref{introduction} suggest that it is inappropriate to attempt to extract
the total mass of each lens galaxy. Keeping all this in mind, we will 
quantify the scaling between GMCF and lens luminosity in the following
simple way.

We define an aperture mass ($M_{260}$)
for each class of lens objects by fitting its
GMCF to a projected singular isothermal sphere (SIS) model. We simply fit
a one parameter model to the density contrast in the aperture:
\begin{equation} \label{SIS}
\Delta\Sigma(R)_{SIS} = \Sigma(R)_{SIS} = { \sigma_v^2 \over 2 G } {1 \over R} 
\end{equation}
where $\sigma_v$ is the central line of sight velocity dispersion. 
This density is then integrated
to find the mass within the aperture. This is not simply the mass associated
with the central galaxy, but should instead be thought of as the normalization
of the GMCF.

As the outer regions of the GMCF are strongly affected by neighbors of the 
lens galaxy,
we restrict this fit to the central 260 \hinv kpc; the inner three bins 
in our GMCF (the normalization A discussed
in \S \ref{gmcf_luminosity} was derived from fits to the entire profile). 
We compare this aperture mass (M$_{260}$) to the lens luminosity
to derive a mass-to-light scaling relation in each of the five bands.

At present the data do not allow one to distinguish between SIS and 
more realistic profiles NFW profiles 
\citep{nfw97}, as both provide 
adequate fits to the observations. NFW profiles decline more rapidly 
at large radii, because the outer slope of NFW profile approaches $R^{-2}$, 
but the error bars are still large there. Moreover, in this region the contribution 
from individual halos becomes small
compared to the contribution from halos with multiple galaxies
or from nearby clustered galaxies in separate halos.
The main difference between SIS and NFW is that in SIS model the 
mass of the galaxy increases with radius more rapidly than in NFW.
If we fix the amplitude of GMCF for the two models 
at 100$h^{-1}$kpc, then we find that at 260$h^{-1}$kpc SIS overestimates 
the mass by a factor of 2 relative to the virial mass of the NFW halo 
(defined as the mass within the radius where the 
mean overdensity is 340 relative to the mean, the value appropriate for the 
cosmological model with $\Omega_m\sim 0.3$ and $\Omega_{\Lambda}=0.7$).
One can therefore interpret the masses in table \ref{gmcftable} as 
the virial masses by taking one half of quoted values.
For example, the best fitted NFW profile to the overall sample gives 
a virial mass of $10^{12}h^{-1}$M$_{\sun}$, with the corresponding virial 
radius of 200$h^{-1}$kpc.

\begin{figure}[t]
\plotone{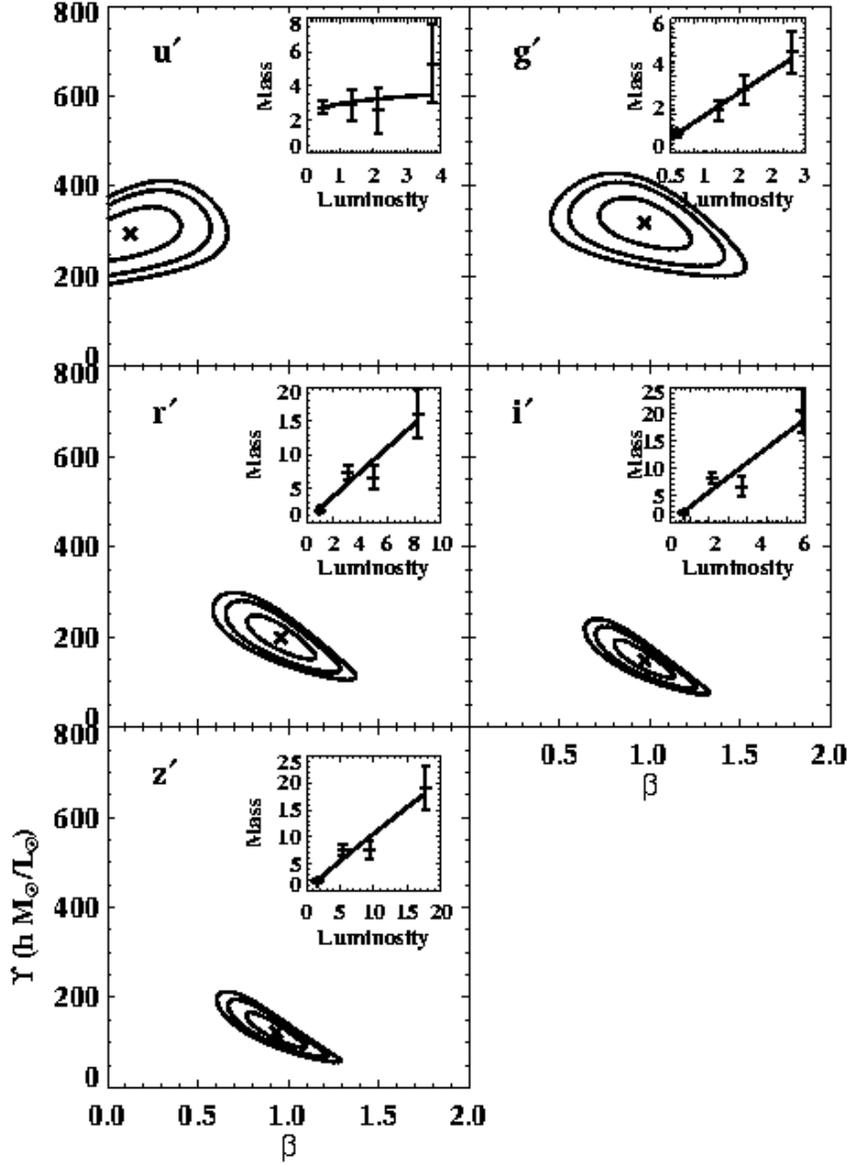}
\figcaption[f18_sm.ps]{
The five panels in this figure summarize the relation between \map\ and
luminosity in each of the five SDSS bands. For each band the small inset 
figure shows this directly. Points in these inset figures are the 
measured \map\ and mean luminosity of galaxies in four luminosity bins.
The line in these inset figures shows the best fit to a power law relation
between \map\ and luminosity of the form: 
$ M_{260} = \Upsilon \times \left(L_{central} /
10^{10} L_{\sun}\right)^{\beta}$. The larger figure shows 68\%, 95\%, and
99\% confidence contours for the fit parameters $\Upsilon$ and
$\beta$.
\label{massvslightfits}
}
\end{figure}

For each color, we have an aperture mass (or GMCF normalization) and mean 
lens luminosity for galaxies 
in each of the four luminosity bins discussed in \S \ref{gmcf_luminosity}.
We fit the data in each color to a power law in the form:
\begin{equation}
M_{260} = \Upsilon \times \left(\frac{L_{central}}
{10^{10} L_{\sun}}\right)^{\beta}
\end{equation}
Results for each of the five colors are summarized in Figure 
\ref{massvslightfits} and Table \ref{lumbintable}. In this figure, the small 
inset plots display the aperture mass in units of $10^{12}$ 
\msun\ vs. the mean luminosity of the central galaxies in units of 
$10^{10}$ \lsun. The strong dependence of \map\ on luminosity in the
\gp, \rp, \ip, and \zp\ bands is apparent. This result was anticipated 
by predictions
based on semianalytic galaxy formation models \citep{guz01}. 
\map\ shows little dependence on \up.
The larger plots show $\chi^{2}$ contours for the fits to 
the parameters $\Upsilon$ and
$\beta$.

A striking result of this analysis is that the relation between 
\map\ and luminosity
is consistent with linear in all bands but \up. Where this is true, it is 
sensible to describe the parameter $\Upsilon$ as a mass-to-light ratio
\mlap. For \up, where mass is clearly not proportional to luminosity, 
$\Upsilon$ can only be considered as the normalization of this very weak
power law fit. This linear scaling (already hinted at by \citet{smi00})
is in contrast to the naive
expectation from the Tully-Fisher or Faber-Jackson relation, which would
predict $\beta \approx 0.5$ for SIS. 

More realistic models based on NFW 
profiles can be made consistent with $\beta \sim 1$ \citep{mmw}. However, 
the Tully-Fisher relation does not lead to a unique mass-luminosity 
relation at the virial radius, because large variations in 
the deduced virial velocity of the halo at a given 
observed rotation velocity can conspire to give approximately the 
same Tully-Fisher relation \citep{navarro01}. 
Galaxy-galaxy lensing provides a much more direct measurement of 
total galaxy masses at a given luminosity and our results suggest that 
mass at 260$h^{-1}$kpc linearly scales with light at these luminosities
(around $L_*$). 

We caution that one must be careful in interpreting 
 \mlap\ as the mass to
light ratio of the central galaxy. On average, it is contaminated by 
contributions from
neighboring objects at the 10\% level (see \S \ref{profmeas} for a description
of this estimate). It is also only an aperture measurement, which does 
not include the total mass of the central galaxy (which may be a poorly 
defined concept anyways). 
The value obtained for such an aperture mass-to-light
ratio has long been known to be dependent on aperture
\citep{ost74,zar99,bah00}. 
If one is interested in virial mass to light 
ratios it is best to fit the observations to NFW profiles directly.
Since the radii where the data are most sensitive to 
are close to the virial radius of a $\times 10^{12}h^{-1}M_{\sun}$ galaxy 
(200$h^{-1}$kpc) one finds that the virial mass is also approximately 
linearly proportional to light. 
\mlap\ is also not a 
mass-to-light ratio which can be simply multiplied by the
luminosity density to determine the cosmic mass density by Oort's method.
This is because we only determine \mlap\ over a narrow range of 
luminosities and there is no reason why the same value should extend to 
all the galaxies that contribute to the luminosity function.
We plan to investigate mass-to-light ratios 
by combining measurements of the galaxy-mass correlation function with measurements
of the galaxy-luminosity correlation function. This analysis will be presented
in a companion paper.

\subsection{Aperture mass-to-light scalings and morphology}

We showed in \S \ref{gmcf_type} that the GMCF for ellipticals is significantly
stronger than that for spirals. Converting the density contrast to an 
aperture mass, we find that \map\ for ellipticals is a factor of 2.7 larger 
than that for spirals. We compare this mass to the mean luminosity of the lens
samples in each of the SDSS bandpasses to obtain a value for \mlap.
We lack the signal to noise to confirm that mass is proportional to light
in each of these samples.

\begin{figure}[t]
\plotone{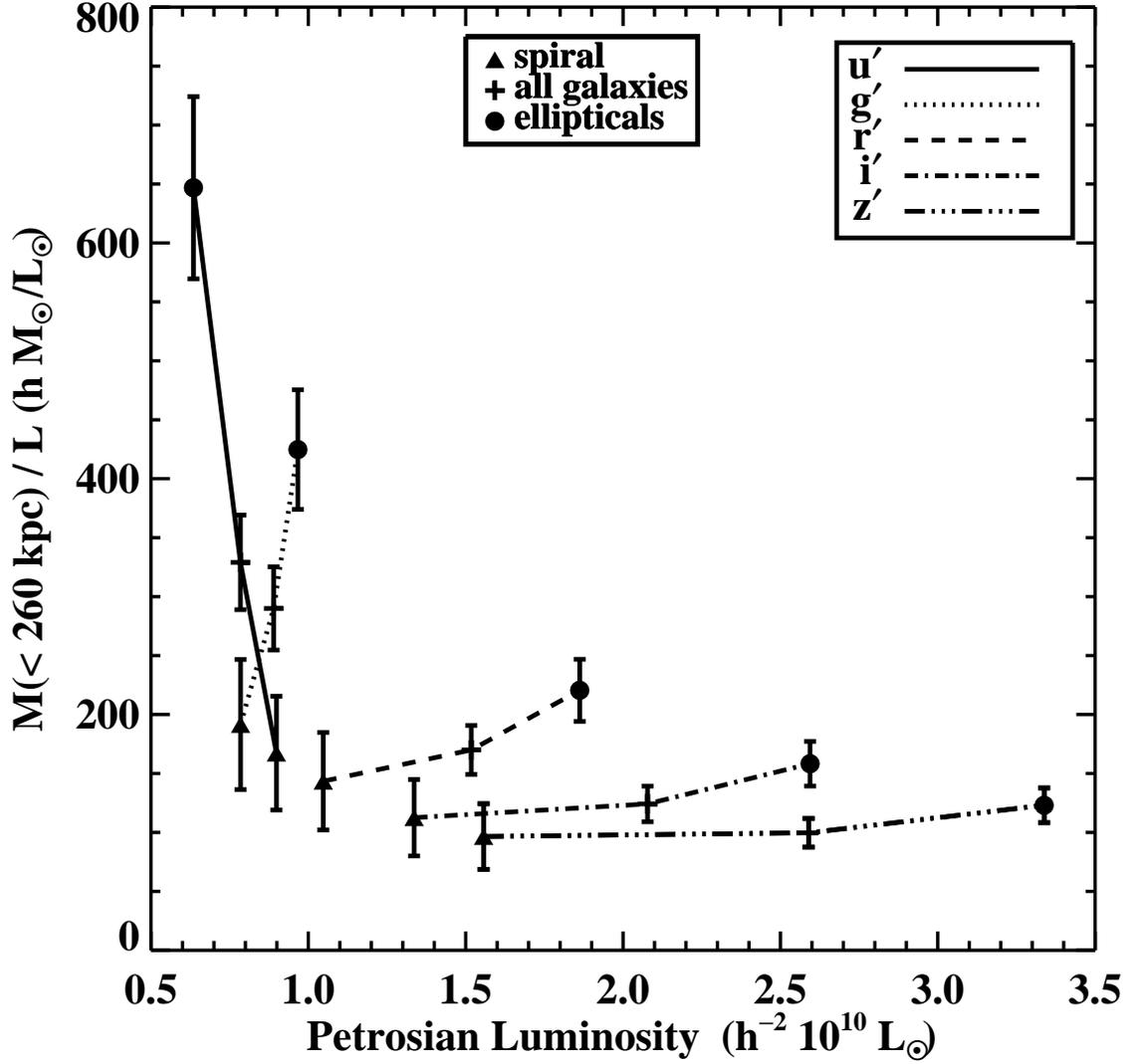}
\figcaption[f19.eps]{
This figure shows the mass-to-light ratios \mlap\ for various galaxy types in 
each of the SDSS bands. Filled triangles represent the spiral sample, 
filled circles the elliptical sample, and plus signs all galaxies.
While \mlap\ ratios are very type dependent in \up\ and \gp, they are 
essentially type independent in the redder \rp, \ip, and \zp\ bands.
\label{mass2lighttype}
}
\end{figure}

Results are shown in Figure \ref{mass2lighttype}. We find that \mlap\ is 
nearly  a factor of four larger for ellipticals than spirals in \up, and 
about a factor of 2.5 larger in \gp. In \rp, \ip, and \zp\, however, the
spiral and elliptical samples show values for \mlap\ which are consistent with
each other, and with the overall sample. We note that the Petrosian 
luminosities used by the SDSS systematically underestimate the luminosity of 
elliptical galaxies relative to spirals by about 20\% relative to spirals
(see \S \ref{luminosity} and \citet{bla01} for details). Accounting for
this makes the \mlap\ aggreement in the red bands still closer.

This reinforces the notion, already suggested in \S \ref{m2l}, that 
the relationship between mass and light is particularly simple when observed
in red bands.

\subsection{Constraining the mean galaxy halo profile from the galaxy-mass 
correlation function} \label{profmeas}

Because the density of lens galaxies on the sky is relatively large, 
and especially because lens galaxies are themselves clustered, 
understanding how much of the projected density we measure is primarily 
associated 
with each central galaxy is complex \citep{fis00,sel00}. The measured density
contrast includes contributions from the central galaxy as well as 
the neighboring galaxies within the aperture. For relatively small radii 
the projected density is dominated by the central galaxy. As the aperture we 
study increases, the average contribution of neighbors also increases. 

There are three kinds of effects. First, there are neighbor contributions which
arise because of the clustering of galaxies in our lens sample. These 
are galaxies which are comparable in luminosity (and hence probably in mass)
to one another. 
A second kind of neighboring object exists as well; those neighbors which are
substantially {\it less} luminous than our lens sample. Dealing with these
galaxies is a definitional problem which lies at the core of this measurement.
Finally, the galaxies can also belong to larger structures, 
such as groups and clusters. There can be mass associated with these 
structures which is not correlated with individual galaxies.

If we wanted to extract the ``total'' mass of galaxies from these measurements,
we would have to account for all possible lens objects. At some point (at 
some luminosity) we would have to begin considering faint neighbors 
as parts of the central lens. As there is no physically motivated scale on
which to do this, we must recognize this as a feature of these measurements. 
Similarly, to include the larger structures we would first need to know 
their positions and masses, which again is not possible in the absence of 
additional information such as X-ray emission. 
As a result most of our analyses treat the GMCF as the 
fundamental measurement. By explicitly including all the mass correlated 
with galaxy locations we avoid the conundrum of how to generate a complete 
sample of lens objects. 

We present here a simple analysis designed to assess the importance of
the first kind of neighbors; those which are luminous enough to be included
in our lens sample. The goal is not to extract isolated ``total'' masses for
lenses, but simply to give the reader an understanding of the effect of lens
clustering on the observed GMCF. 

\begin{figure}[t]
\plotone{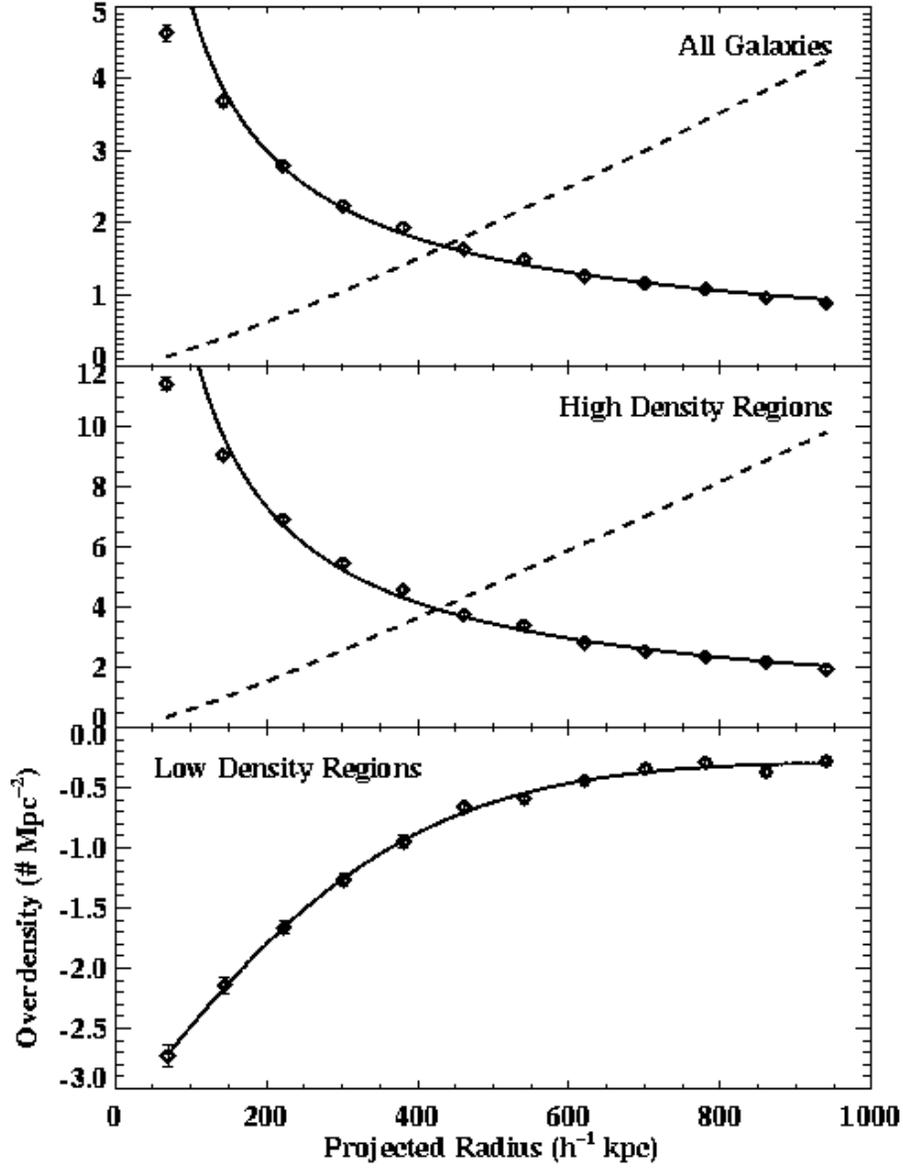}
\figcaption[f20_sm.ps]{
Overdensity of bright galaxies around the lens galaxy sample as compared
to random points.  The overdensity for all lens galaxies, lens galaxies
in high density regions and galaxies in low density regions is shown
from top to bottom. For the top two plots, the solod line is the
best fit power law and the dashed line is the measured
cumulative excess number of neighbors. The inner bin deviates from a 
power law due to undeblended neighbors of the central galaxy.
In the bottom plot, the solid line is the
best fit negative gaussian.
\label{wtheta_all}
}
\end{figure}

We begin by determining the extent of clustering among our lens galaxies.
Figure \ref{wtheta_all} shows the overdensity $\phi(R)$ of neighboring 
bright galaxies
(\rs $\le$ 17.8) around our foreground lens sample (symbols with error
bars).  The dashed line is the cumulative number of bright neighbors.
For this sample, there is on average about 1 neighbor within $275 h^{-1}$ kpc.
Each of these neighbors contributes to the measured density contrast.
The solid line is the best-fitting power law, 
\begin{equation}
\phi(R) =  \left( \frac{R}{0.94\textrm{Mpc}} \right)^{-0.74} 
\textrm{Mpc}^{-2}
\end{equation}
The data deviate from a power law in the inner bin. We believe this
is due to undeblended neighbors of the central
galaxy; see \citet{scr01} for more details.
We note that this is not quite the same as measuring the overdensity
of lens galaxies around lens galaxies. For our lens sample we require
an SDSS spectroscopic redshift. There are some galaxies which are bright
enough to be targeted for SDSS spectroscopy which do not as yet have 
spectra. While we cannot include these galaxies in our lens sample (it
would be impossible to rescale them correctly), we do include them in 
our measurement of the clustering of lens galaxies.

We follow the method of \cite{fis00} to account for the presence of the
neighboring galaxies. We assume that all the galaxies in the sample
have the same halo profile, $\Sigma_g(R)$. We further assume
the density of neighbors is given by the overdensity shown in 
Figure \ref{wtheta_all}. The mean projected mass density of the
neighbors can then be written as 
\begin{equation} \label{neighconv}
\Sigma(R)_{neigh} = [\Sigma_g(R^{\prime}) * \phi(R^{\prime}) ](R)
\end{equation}
and the total mean density contrast can be written as
\begin{equation} \label{tdcont}
\Delta \Sigma(R) = \Delta \Sigma_g(R)\ + \Delta \Sigma_{neigh}(R).
\end{equation}
Note that neighboring galaxies do not contribute linearly to the density 
contrast because they lie at the edge of the aperture.
Because fainter galaxies are not counted in our overdensity $\phi$, their
masses are implicitly assigned to the ``mean'' mass profile.

To perform the deconvolution of equation \ref{tdcont} and recover
the halo density profile, we would need to extrapolate both the
overdensity of neighbors and the GMCF to infinity.
We instead take a simpler approach, which only requires extrapolating
the overdensity. Following \cite{bra96} and \cite{fis00}, we 
assume the galaxy profile is represented by a simple model:
\begin{equation}
\rho(r) = \frac{\sigma_v^2 s^2}{ \pi G r^2 \left( r^2 + s^2 \right) }
\end{equation}
where $\sigma_v$ is the central velocity dispersion for $r \ll s$ and 
s is an outer scale length. For $r \ll s$, this profile resembles an
isothermal ($\propto r^{-2}$), and for $r \gg s$ the 
profile cuts off quickly as $r^{-4}$.
Integrating along the line of sight, we get the projected density profile
\begin{equation} \label{projdensmod}
\Sigma_g(R) = \frac{\sigma_v^2}{2 G R}
            \left( 1 - \frac{R}{\sqrt{R^2 + s^2} } \right).
\end{equation}
We see that this is the same as the SIS in equation \ref{SIS} with a cutoff 
radius s.
This model has a finite mass, with $M_{tot} \varpropto s$.

Taking $\phi(R) = \left(R/0.94\textrm{Mpc}\right)^{-0.74}$, we convolve the 
model density with $\phi(R)$ 
as in equation \ref{neighconv}
and construct the model density contrast as in equation \ref{tdcont}. We then
perform $\chi^2$ fitting of this model against the data shown in figure
\ref{all_lens_profile} to extract the best-fit 
galaxy parameters $\sigma_v$ and s.

\begin{figure}[t]
\plotone{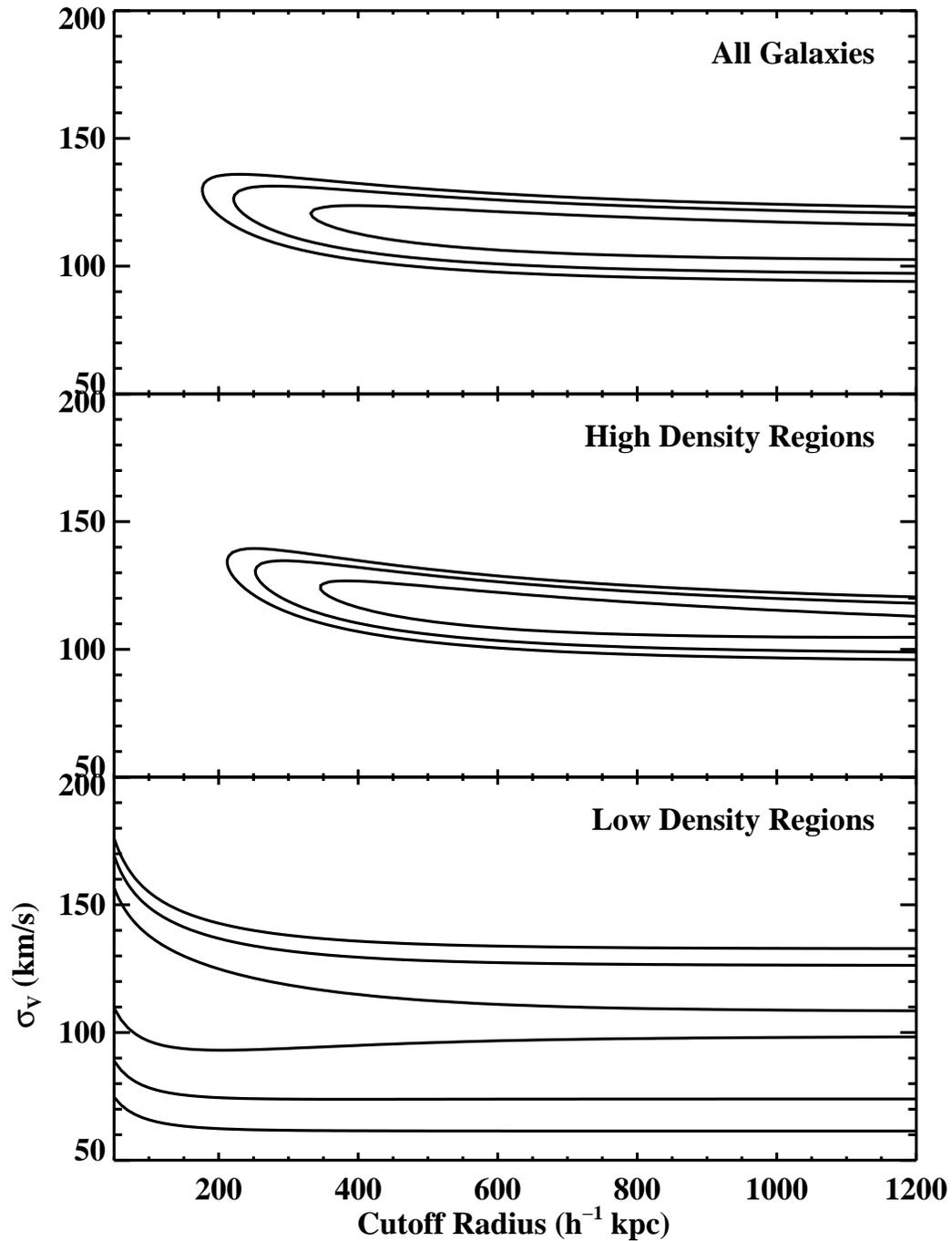}
\figcaption[f21.eps]{
$\chi^2$ contours from fits to truncated isothermal spheres for all galaxies,
galaxies located in regions of high local galaxy density, and galaxies
in regions of low local densiy, shown from top to bottom, respectively.
Fits include
the contribution to the projected density contrast due to neighboring
galaxies. For the average galaxy, the velocity dispersion $\sigma_v$ is 
well-constrained,
but only a lower bound can be placed on the outer scale. The fits for the low
and high density samples is consistent with that of the average galaxy.
\label{chisq_cont_all} 
} 
\end{figure}

The top plot in Figure \ref{chisq_cont_all} shows the 68\%, 95\%, and 99\% confidence regions
for these fits.
We have used the full covariance matrix to estimate
the $\chi^2$ statistic.
As in \cite{fis00}, we can only put a lower limit on the outer scale length,
s $\ge 230 h^{-1}$ kpc (95\% conf.).
The velocity dispersion is well constrained, however, to be in the range
100-130 km s$^{-1}$ (95\% conf.), with a best-fit value of 
$\sigma_v = 113$ km s$^{-1}$ ($v_c = 160$ km s$^{-1}$). 

It is worth 
commenting on the comparison of this result to the results in \citet{fis00}. 
The size of the lens samples are comparable ($\sim$28,000 there vs. 
$\sim$31,000 here), and hence
the sensitivity to cutoff radius is similar. The difference between these 
lens samples is that the \citet{fis00} sample was magnitude selected 
from \rs=16-18, and this sample is selected(by the standard SDSS 
spectroscopic survey selection) as all galaxies with \rs $\leq$ 17.6. The 
effect is to focus on a less luminous, and
hence less massive, lens sample. In addition, we now know from SDSS 
spectroscopy that the mean redshift estimated for the \citet{fis00} 
lens sample was overestimated by about 35\%. Correcting this would 
reduce the best fit $\sigma_v$ for that sample by about the same 
factor. The results here are in agreement with our earlier results. Since
they are based on lens samples with spectroscopic redshifts these
results are substantially more robust.

The best fit total density contrast and the corresponding central galaxy
are overplotted in the top plot of Figure \ref{all_dense_profile}. As 
expected, the 
density contrast is completely dominated by the central galaxy for 
small radii, while the contribution from neighbors becomes increasingly 
important for larger radii. At about 950 \hinv
kpc the contribution from neighbors 
dominates. It is worth noting that, although there is on average 1 neighbor
within 260 \hinv kpc, the aperture used in \S \ref{m2l}, the net contribution
of neighbors to the {\it density contrast} within that radius 
is $\lesssim$ 10\%.
For the average galaxy, the contribution of neighboring galaxies is not a 
dominant factor in the \mlap\ measurements described in \S \ref{m2l}.

There are in the literature other approaches to extracting the ``isolated
mean galaxy'' from lensing data of this kind. \citet{sch97} have presented
a method in which a model galaxy, with certain luminosity scaling relations,
is fit to the entire set of source and lens galaxies. An intellectually
similar, but technically quite different approach has been developed by
\citet{joh01}. While these techniques present an interesting alternative 
to the more direct measurement of the GMCF, they require knowledge of all lens
objects contributing to the shape of each source galaxy. This information is
only available for objects far from the edges of the observed region. These
SDSS data, taken in long strips, have little area far from the edges. As a 
result, we will pursue this alternative analysis when a number of 
contiguous SDSS stripes are available for lensing studies.

\subsection{Tests of galaxy profile determination}

If the deconvolution method described in \S \ref{profmeas} is properly
removing the effects of neighbors, we should be able to choose galaxies
from different environments, but with the same luminosity, and recover  
consistent halo parameters. 
We use the Voronoi tesselation of the galaxy positions described in
\S \ref{gmcf_environment} to select samples of galaxies from high and 
low density regions. The galaxies from high and low density regions have
\is\ luminosity that is similar to the mean luminosity of the entire lens
sample (2.1, 2.0, and 2.0$\times 10^{10} L_{\sun}$ respectively). The 
overdensity of neighbors for the high and low density samples
is measured in the same manner desribed in \S \ref{profmeas}.
Results are shown in Figure \ref{wtheta_all}.  
The galaxies selected to have high local density do have a higher $\phi$
by about a factor of two at every radius, and the best-fit power
law is
\begin{equation}
\phi_{high}(R) =  \left( \frac{R}{1.74\textrm{Mpc}} \right)^{-0.82}
\textrm{Mpc}^{-2}.
\end{equation}
For the galaxies selected to have lower local density, we find that 
the amplitude $\phi_{low}$ is actually negative, indicating that
these galaxies tend to live in underdense regions. We find that $\phi_{low}$
does not fit well to a power law. We represent it instead as a Gaussian
with scale length 400 \hinv kpc, and use this to do the deconvolution.

\begin{figure}[t]
\plotone{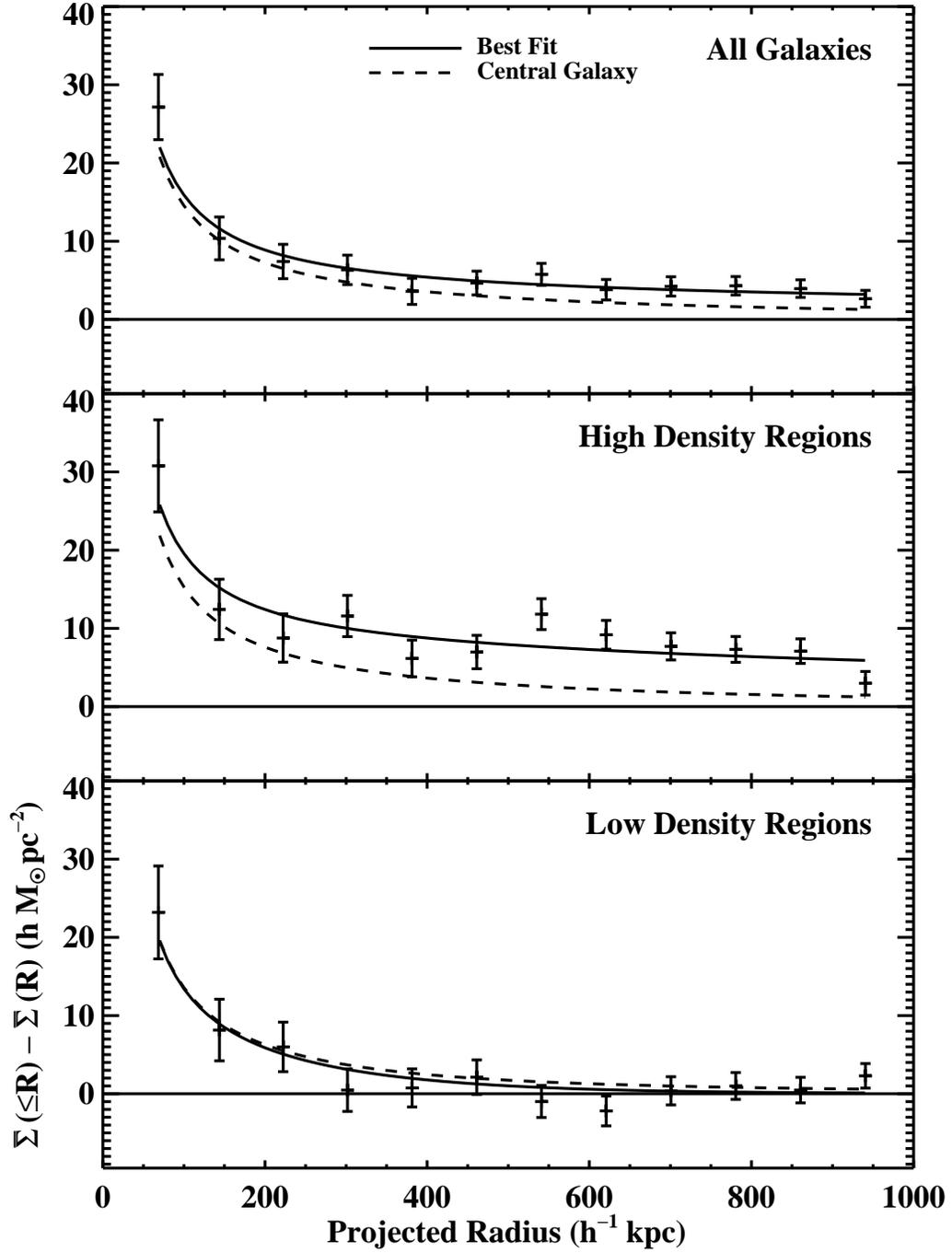}
\epsscale{0.7}
\figcaption[f22.ps]{Same as figure \ref{all_lens_profile} 
except this plot shows the combined density contrast for all galaxies, galaxies
in high density regions, and galaxies in low density regions, shown from top
to bottom, respectively. 
The solid line is the best-fitting density contrast from a model which includes
the central galaxy and neighbors. The dashed line shows only the contribution
from the central galaxy.
The contribution from neighboring galaxies varies substantially for
galaxies in high and low density regions, but
the mean central galaxy profile is consistent with that of the average
galaxy. This agreement, and the fact that the sub-samples have 
similar luminosity, suggests that the deconvolution method is accurately 
accounting for the effects of comparably bright neighbors in the projected 
density contrast.
\label{all_dense_profile}}
\end{figure}

The density contrast measured around these sub-samples is shown in Figure
\ref{all_dense_profile}. Indeed, the signal at large radii flattens
out as expected for galaxies in higher density regions, and the signal from
underdense regions drops more rapidly to zero.

Repeating the deconvolution analysis of \S \ref{profmeas} on the high density
lens sample, we find the confidence regions for $\sigma_v$ and s shown
in the middle plot of figure \ref{chisq_cont_all}. The allowed values for the velocity dispersion
agree quite well with those found for the entire lens sample, 
100-135 km s$^{-1}$ (95\% conf). We still find only a lower bound on
the cutoff radius, $s \ge 260 h^{-1}$ kpc (95\% conf.). The best-fit 
model and central
galaxy are overplotted on the middle figure \ref{all_dense_profile}, demonstrating
the large contribution to the density contrast made by the neighboring
galaxies.

For the low-density sample, we find that the void-like regions surrounding
these galaxies actually {\it reduces} the lensing signal, and density
contrast must be added to the measured signal to reconstruct the central
galaxy. Although we cannot constrain the outer scale at all, we find that 
the best-fit velocity dispersion is still consistent with that 
inferred for the average galaxy, 75-170 km s$^{-1}$ (95\% conf.), 
with a best-fit value of 125 km s$^{-1}$.

This test demonstrates that we are able to recover similar halo parameters
for galaxies of similar luminosity that come from very different environments.
This suggests that we have a reasonable understanding of the effects of 
bright neighboring galaxies on the density constrast.

\section{Summary and Discussion} \label{discussion}

Our understanding of structure formation in the universe has evolved 
substantially over the last thirty years. We no longer expect galaxies 
to form as discrete, isolated objects. Nor is the formation of a galaxy 
simply a series of mergers of discrete, otherwise isolated objects. 
Instead, luminous galaxies are roughly discrete 
tracers of mass, embedded in a smoothly varying dark matter environment. 
Over time, 
the very smooth matter distribution of the early universe becomes more 
and more inhomogeneous. Despite this, galaxies are today embedded in dark 
matter halos which overlap to a degree which invalidates discussion of them 
as discrete objects.

We have presented measurements of a galaxy-mass correlation function and 
its dependence on a variety of measured galaxy properties. We measured the 
dependence of the GMCF on galaxy luminosity. While the shape of the GMCF is
little affected by luminosity, the amplitude can vary strongly.
The GMCF has little relation to \up\ luminosity, reflecting the fact that 
much of this 
luminosity is derived from localized, short lived episodes of star 
formation. The GMCF is strongly dependent on the luminosity of 
galaxies in the \gp, \rp, \ip, and \zp\ bands. 
The luminosity in these bands is 
dominated by low mass stars, and hence reflects the integrated star formation 
history of the galaxy.

We have examined the relationship between the GMCF and the morphology of 
galaxies. The GMCF is much stronger for elliptical 
galaxies than it is for spirals. Ellipticals typically reside in 
halos which are more massive than spirals.

We examined the relationship between GMCF and galaxy environment. We found 
that the amplitude of the GMCF near the center is very similar for 
galaxies of the same luminosity drawn 
from overdense and underdense regions, yet the shape of the GMCF
at large radii is quite different. This result confirms the notion that 
the GMCF in the central regions, out to perhaps 300 \hinv kpc, is dominated 
by mass clearly associated with the central galaxy. The GMCF at large 
radius, by contrast, is primarily effected by the presence of neighboring 
objects.

In a straightforward effort to quantify the scaling of the
GMCF with central galaxy luminosity ($L_{central}$) we derived 
aperture masses for various
classes of lenses based on fits to the inner 260 \hinv kpc of their 
GMCF ($M_{260}$). Comparison of $M_{260}$ to the mean $L_{central}$ confirms
a linear scaling in the \gp, \rp, \ip, and \zp\ bandpasses. 
These measurements of $M_{260}$
also allowed us to probe quantitatively the \mlap\ relations for 
spiral and elliptical galaxies. Despite the substantial 
difference between the GMCF for
spirals and ellipticals, there is little difference in the relationship 
between their GMCF and luminosities in red bands. Since spiral and 
elliptical galaxies occupy substantially different average local 
environments, the consistency of these GMCF/luminosity relations reinforces 
the notion that the GMCF measured within 260 \hinv kpc is primarily 
associated with the central object.

It is worth noting that many efforts to understand 
the relation between galaxy mass and light have utilized measurements of 
blue light (intermediate between the SDSS \up\ and \gp\ bands). Our 
results suggest that any such attempt is problematic. 
They cast doubt on attempts to explain the
Tully-Fisher relation in B-band with the assumption that 
rotational velocity is a constant fraction of circular velocity at 
virial radius \citep{mmw}. In the red bands the situation 
is significantly simpler, and our data are consistent with the assumption
that light traces mass on halo scales, independent of luminosity and 
morphological type. Future studies of galaxy mass and its relation to 
luminosity should concentrate on luminosities measured in red bands.

Finally, we made an effort to understand the effect of galaxy clustering on
naive interpretations of these measurements. By deconvolving the GMCF and
the lens-lens correlation function we obtain rough estimates of the relative
contributions of lens galaxies and comparably luminous neighbors. At
the 260 \hinv kpc radius of our \mlap\ measurements we estimate
that the contamination from neighbors is typically 10\%.
No evidence for a cutoff in the deconvolved GMCF is 
seen. We place 95\% confidence limits on such a cutoff at 230 \hinv kpc 
within the context of SIS models. Current data are not sufficiently 
accurate to distinguish between SIS and NFW profiles and the latter 
also provide a good fit to the data. 

It appears that the integrated star formation history of a galaxy (as
reflected in its \ip\ luminosity for example) is 
closely correlated with the depth of the dark matter halo
in which the galaxy formed. This statement seems to be true independent of 
the luminosity or morphology of the galaxy.
Two factors are required for this correlation to hold. First, 
the fraction of baryonic to total mass on halo scales must be roughly 
constant. If this baryon fraction is constant, the amount of material 
available to 
form stars will be related to the halo mass. That the baryon fraction 
should be roughly constant is not too surprising, as the regions from which 
these halos are drawn are quite large. Altering the baryon fraction 
substantially on these scales would require a mechanism for either 
separating baryons from dark matter before halo potentials form, or for 
moving baryons from one potential well to another.

The second requirement for a close relation between mass and luminosity is 
a roughly constant efficiency for star formation in galaxies of different 
sizes. This efficiency is clearly not constant when measured on short 
timescales. This is why the GMCF is so weakly related to the \up\ 
luminosity of galaxies. The luminosity of galaxies in the red bands, 
however, reflects the integrated star formation history of the galaxies. 
The direct relation between the GMCF and these red luminosities suggests 
that this integrated star formation efficiency is, in fact, relatively 
constant across a range of galaxies.

These measurements are the first of many which can be made with existing 
and future SDSS data. They provide a direct comparison between the luminous 
properties of objects and the projected surface density of the dark matter 
environment in which they reside. These weak lensing measurements are 
particularly powerful 
for their ability to provide a consistent approach to the study of objects 
with a wide range of properties. They can be used in an identical way with 
galaxies of all types, luminosities, and environments.

\section{Future prospects} \label{future}

While the studies presented here go substantially beyond previous 
galaxy-galaxy lensing studies, they represent only the beginning of what 
can be done with SDSS data. All the studies described here are based on 
SDSS commissioning data which does not meet survey data quality 
requirements. The survey data now being 
taken have substantially better image quality and PSF stability. The 
improved image data allow us to better measure galaxy shapes 
and to include more galaxies in our source catalog. More important, 
the data presented here are drawn from only about 4\% of the final SDSS 
survey area. The full data set will allow repetition of all these 
measurements with more than 5 times the signal to noise.

The enhanced statistical power provided by the full SDSS can be used in an 
alternative way as well. We can divide the universe of galaxies into as 
many as twenty-five subsets, and obtain for each a GMCF with S/N comparable 
to those presented here. This will allow us to study scaling relations 
between the GMCF and a wide range of galaxy properties; including scale 
length, local environment, evidence of recent mergers, stellar population 
and age estimates, etc.

Additional analysis of galaxy halo concentrations is possible. SDSS spectra 
are of sufficiently high quality to provide measured velocity dispersions 
for essentially every early type galaxy \citep{bernardi01}. Typical errors 
in $\sigma_v$ are $<$15 km/s. These velocity dispersions 
provide a dynamical measure of the mass in the central 
$\sim$10 \hinv kpc of these galaxies. 
Comparison of these measurements to the GMCF 
measurements on 250 \hinv kpc scales will provide a direct probe of 
variations in the concentration of galaxy halos. This analysis is underway 
now, using a sample of more than 10,000 early type galaxies with measured 
$\sigma_v$.

As mentioned above, there are other analysis techniques which can be 
applied to this data \citep{sch97,joh01}. They will be much more effective 
when the SDSS has completed imaging and spectroscopy of a region with a 
less extreme axis ratio.

A very important project will be direct comparison of these results to 
measurements  conducted in the same manner in the results of N-body 
simulations. 
Some initial work based on the GIF \citep{kau99} semianalytic simulations has 
been done by \citet{guz01} and \citet{ste01}. 
Another promising approach is to use 
halo models to quantify galaxy-galaxy lensing \citep{sel00}. The 
ultimate outcome of such analyses will be the galaxy luminosity function as a 
function of halo mass.
We expect to substantially expand this 
work in the future, providing essential feedback to the models of 
galaxy formation.

Extension of this approach to studies of other kinds of `objects' is also 
possible. In addition to the galaxy-mass correlation function described 
here, we can measure the group-mass correlation function, the cluster 
mass correlation function, even a void-mass correlation function. All 
that is required is an input catalog of object locations and redshifts. A 
simple example demonstrating our ability to measure the cluster-mass 
correlation function appeared in \citet{she01}.

Weak lensing within the SDSS provides a powerful new tool for probing the 
relationship between luminous objects and the dark matter environment in 
which they reside. This is particularly timely as we enter an era of 
precision cosmology, in which the cosmological parameters are well 
constrained, and these details of galaxy bias can no longer be ignored. 
SDSS lensing measurements provide strong constraints on the total 
correlation between objects and mass, and hence provide simple, direct 
constraints to N-body simulation results.



\acknowledgments

The Sloan Digital Sky Survey (SDSS) is a joint project of The 
University of Chicago, Fermilab, the Institute for Advanced Study, 
the Japan Participation Group, The Johns Hopkins University, the
Max-Planck-Institute for Astronomy (MPIA), the Max-Planck-Institute 
for Astrophysics (MPA), New Mexico State University, Princeton 
University, the United States Naval Observatory, and the University 
of Washington. Apache Point Observatory, site of the SDSS
telescopes, is operated by the Astrophysical Research Consortium (ARC). 
Funding for the project has been provided by the Alfred P. Sloan 
Foundation, the SDSS member institutions, the National Aeronautics 
and Space Administration, the National Science Foundation, the U.S. 
Department of Energy, the Japanese Monbukagakusho, and the Max Planck 
Society. The SDSS Web site is http://www.sdss.org/. 
Timothy McKay and Erin Sheldon acknowledge support from NSF PECASE grant 
AST 9708232. The authors also gratefully acknowledge continued discussions 
of all aspects of weak lensing with Gary Bernstein and his students.





\clearpage



\clearpage

\clearpage

\clearpage

\clearpage

\clearpage

\clearpage


\begin{deluxetable}{ccccc}
\rotate
\tabletypesize{\small}
\tablecaption{Lensing Data and Power Law Fits for the Galaxy-Mass Correlation
Function \label{gmcftable} }
\tablewidth{0pt}
\tablehead{
\colhead{Sample \tablenotemark{a}} &
\colhead{N$_{Lenses}$ \tablenotemark{b}} &
\colhead{$\langle\Delta\Sigma_+(20 \le R \le 980$ kpc$)\rangle$ \tablenotemark{c}} &
\colhead{A \tablenotemark{d}} &
\colhead{$\alpha$ \tablenotemark{e}} \\
&
&
\colhead{h M$_{\sun}$ pc$^{-2}$} &
\colhead{h M$_{\sun}$ pc$^{-2}$} &
}
\startdata
All \gp               & 31039 & 2.77 $\pm$ 0.65 & 1.16$^{+0.72}_{-0.52}$ & 1.19$^{+0.26}_{-0.21}$ \\
\smallskip
All \rp               & 31174 & 4.73 $\pm$ 0.51 & 3.11$^{+0.86}_{-0.78}$ & 0.65$^{+0.21}_{-0.19}$ \\
\smallskip
All \ip               & 31203 & 5.70 $\pm$ 0.53 & 3.63$^{+0.86}_{-0.78}$ & 0.69$^{+0.17}_{-0.16}$ \\
\smallskip
All combined          &  -    & 4.59 $\pm$ 0.42 & 2.55$^{+0.68}_{-0.60}$ & 0.82$^{+0.17}_{-0.16}$ \\
\smallskip
Spirals combined (\gp/\rp/\ip)      & 13806/13882/13904 & 2.28 $\pm$ 0.58 & 1.14$^{+1.04}_{-0.74}$ & 0.88$^{+0.57}_{-0.51}$ \\
\smallskip
Ellipticals combined (\gp/\rp/\ip) & 14097/14156/14157 & 7.52 $\pm$ 0.65 & 4.33$^{+1.04}_{-0.96}$ & 0.81$^{+0.15}_{-0.15}$  \\ 
\smallskip
Low Density Regions (\gp/\rp/\ip)  & 15114/15185/15197 & 1.18 $\pm$ 0.62 & 0.25$^{+0.53}_{-0.23}$ & 1.72$^{+0.51}_{-1.06}$  \\ 
\smallskip
High Density Regions (\gp/\rp/\ip) & 15568/15631/15642 & 7.82 $\pm$ 0.59 & 5.45$^{+1.09}_{-1.02}$ & 0.56$^{+0.17}_{-0.15}$  \\ 
\enddata

\tablenotetext{a}{``All'' means all galaxies in the sample. Bandpasses are those in which source galaxy shapes were measured}
\tablenotetext{b}{Number of lenses which passed lensing cuts}
\tablenotetext{c}{$\Delta\Sigma_+ = \overline{\Sigma}(\leq R) - \overline{\Sigma}(R)$}
\tablenotetext{d}{Normalization of best-fit power law; 
$\Delta\Sigma_{+} = A \left(\textrm{R}/\textrm{1 Mpc}\right)^{-\alpha}$}
\tablenotetext{e}{Power law index of best-fit power law to GMCF}
\end{deluxetable}


\begin{deluxetable}{cccccc}
\rotate
\tabletypesize{\small}
\tablecaption{Lensing Data and Model Fits for Luminosity Bins \label{lumbintable}}
\tablewidth{0pt}
\tablehead{
\colhead{Bandpass \tablenotemark{a}} &
\colhead{N$_{Lenses}$ \tablenotemark{b}} &
\colhead{$\langle L_{Cent} \rangle$\tablenotemark{c}} &
\colhead{$\langle\Delta\Sigma_+(20 \le R \le 260$ kpc$)\rangle$\tablenotemark{d}} &
\colhead{$\langle$M(R$\le$260 kpc) $\rangle$\tablenotemark{e}} &
\colhead{$\langle$M($\le$260 kpc)/L$_{Cent}\rangle$} \\
&
\colhead{\gp/\rp/\ip} & 
\colhead{10$^{10}$h$^{-2}$L$_{\sun}$} &
\colhead{h M$_{\sun}$ pc$^{-2}$} &
\colhead{10$^{12}$ h$^{-1}$ M$_{\sun}$} &
\colhead{h M$_{\sun} / $L$_{\sun}$}
}
\
\startdata
\up & 21041/21141/21185 & 0.494 $\pm$ 0.002 & 7.4 $\pm$ 2.5   & 2.7 $\pm$ 0.4  & 498 $\pm$ 66 \\
 -  & 4815/4833/4819    & 1.342 $\pm$ 0.003 & 15.2 $\pm$ 6.1  & 2.9 $\pm$ 0.9  & 207 $\pm$ 64 \\
 -  & 2433/2435/2439    & 2.133 $\pm$ 0.006 & 11.2 $\pm$ 6.7  & 2.6 $\pm$ 1.3  & 118 $\pm$ 62 \\
 -  & 1120/1132/1124    & 3.76  $\pm$ 0.03  & 23.3 $\pm$ 11.4 & 5.3 $\pm$ 2.3  & 138 $\pm$ 59 \\
    & & & & & \\
\gp & 20960/21061/21096 & 0.631 $\pm$ 0.003 & 9.2 $\pm$ 1.8   & 2.1 $\pm$ 0.4  & 291 $\pm$ 49 \\
 -  & 4910/4933/4928    & 1.561 $\pm$ 0.003 & 19.5 $\pm$ 4.9  & 4.5 $\pm$ 1.0  & 259 $\pm$ 57 \\
 -  & 2440/2439/2434    & 2.130 $\pm$ 0.004 & 28.8 $\pm$ 7.4  & 6.6 $\pm$ 1.5  & 292 $\pm$ 66 \\
 -  & 1110/1120/1120    & 3.21  $\pm$ 0.03  & 45.9 $\pm$ 8.4  & 10.5 $\pm$ 2.2 & 301 $\pm$ 63 \\
    & & & & & \\
\rp & 20924/21031/21076 & 0.991 $\pm$ 0.005 & 8.0 $\pm$ 1.8   & 1.8 $\pm$ 0.4  & 166 $\pm$ 32 \\
 -  & 4813/4829/4808    & 3.070 $\pm$ 0.007 & 32.3 $\pm$ 5.0  & 7.4 $\pm$ 1.0  & 228 $\pm$ 31 \\
 -  & 2526/2536/2538    & 4.95 $\pm$ 0.02   & 28.9 $\pm$ 8.6  & 6.6 $\pm$ 1.7  & 126 $\pm$ 33 \\
 -  & 1119/1121/1119    & 8.23 $\pm$ 0.06   & 70.5 $\pm$ 18.8 & 16.1 $\pm$ 3.7 & 187 $\pm$ 43 \\
    & & & & & \\
\ip & 20966/21074/21118 & 1.298 $\pm$ 0.006 & 7.5 $\pm$ 1.8   & 1.7 $\pm$ 0.3  & 121 $\pm$ 25 \\
 -  & 4730/4747/4728    & 4.28 $\pm$ 0.01   & 36.2 $\pm$ 5.0  & 8.3 $\pm$ 1.0  & 183 $\pm$ 22 \\
 -  & 2608/2614/2617    & 7.35 $\pm$ 0.03   & 29.1 $\pm$ 8.6  & 6.6 $\pm$ 1.7  & 86 $\pm$ 22 \\
 -  & 1095/1098/1094    & 13.7 $\pm$ 0.1    & 90.5 $\pm$ 20.4 & 20.7 $\pm$ 4.1 & 148 $\pm$ 29 \\
    & & & & & \\
\zp & 21036/21148/21191 & 1.587 $\pm$ 0.008 & 7.7 $\pm$ 1.8   & 1.8 $\pm$ 0.3  & 103 $\pm$ 20 \\
 -  & 4711/4723/4706    & 5.42 $\pm$ 0.01   & 33.2 $\pm$ 7.5  & 7.6 $\pm$ 1.0  & 133 $\pm$ 18 \\
 -  & 2579/2586/2587    & 9.35 $\pm$ 0.04   & 33.4 $\pm$ 8.7  & 7.6 $\pm$ 1.7  & 77 $\pm$ 18 \\
 -  & 1091/1094/1091    & 17.7 $\pm$ 0.2    & 83.2 $\pm$ 20.3 & 19.0 $\pm$ 4.0 & 104 $\pm$ 22 \\
\enddata

\tablenotetext{a}{Bandpass in which the luminosities were measured}
\tablenotetext{b}{Number of lenses used.
Here \gp, \rp, and \ip\ refer to the images in which the shapes of source
galaxies were measured}
\tablenotetext{c}{Mean luminosity of lens galaxies in this sample and 
bandpass. Mean luminosities are calculated with the same weights used in the
lensing analysis (see \S \ref{combining})}
\tablenotetext{d}{Combined density contrast as measured from sources found in \gp, \rp, and \ip\ imaging data}
\tablenotetext{e}{Based on fits to singular isothermal spheres}

\end{deluxetable}


\begin{deluxetable}{lcccccc}
\tabletypesize{\small}
\tablecaption{Mass and M/L Fits for Different Galaxy Types \label{typem2ltable} }
\tablewidth{0pt}
\tablehead{
\colhead{Galaxy Type} &
\colhead{$\langle$M(R$\le$260 kpc) $\rangle$ \tablenotemark{a}} &
\colhead{$\langle$M/L$_{u^{\prime}} \rangle$  \tablenotemark{b}} &
\colhead{$\langle$M/L$_{g^{\prime}} \rangle$} &
\colhead{$\langle$M/L$_{r^{\prime}} \rangle$} &
\colhead{$\langle$M/L$_{i^{\prime}} \rangle$} &
\colhead{$\langle$M/L$_{z^{\prime}} \rangle$} \\
&
\colhead{10$^{12}$ h$^{-1}$ M$_{\sun}$} &
\colhead{h M$_{\sun} / $L$_{\sun}$} &
\colhead{h M$_{\sun} / $L$_{\sun}$} &
\colhead{h M$_{\sun} / $L$_{\sun}$} &
\colhead{h M$_{\sun} / $L$_{\sun}$} &
\colhead{h M$_{\sun} / $L$_{\sun}$}
}

\startdata
All         & 2.6 $\pm$ 0.3 & 329 $\pm$ 40 & 290 $\pm$ 35 & 170 $\pm$ 21 & 124 $\pm$ 15 & 100 $\pm$ 12 \\
Spirals     & 1.5 $\pm$ 0.4 & 167 $\pm$ 48 & 192 $\pm$ 55 & 143 $\pm$ 41 & 112 $\pm$ 32 & 97 $\pm$ 28 \\
Ellipticals & 4.1 $\pm$ 0.5 & 647 $\pm$ 77 & 425 $\pm$ 51 & 221 $\pm$ 26 & 158 $\pm$ 19 & 123 $\pm$ 14 \\
\enddata

\tablenotetext{a}{Based on fits to singular isothermal spheres}
\tablenotetext{b}{M = M(R$\le$260 kpc)}
\end{deluxetable}

\end{document}